\documentclass[english,12pt]{article}
\usepackage[T1]{fontenc}
\usepackage{geometry}
\usepackage{color}
\usepackage{babel}
\usepackage{booktabs}
\usepackage{mathtools}
\usepackage{amsmath}
\usepackage{amsthm}
\usepackage{amssymb}
\usepackage{graphicx}
\usepackage{setspace}
\usepackage[authoryear]{natbib}
\usepackage[unicode=true,
 bookmarks=true,bookmarksnumbered=true,bookmarksopen=false,
 breaklinks=false,pdfborder={0 0 0},pdfborderstyle={},backref=false,colorlinks=true]
 {hyperref}

\makeatletter

\providecommand{\tabularnewline}{\\}

  \theoremstyle{definition}
  \newtheorem{defn}{\protect\definitionname}
\theoremstyle{plain}
\newtheorem{thm}{\protect\theoremname}
  \theoremstyle{remark}
  \newtheorem{rem}{\protect\remarkname}

\setcitestyle{round}

\date{}

\makeatletter
\renewcommand\paragraph{\@startsection{paragraph}{4}{\z@}%
    {1.25ex \@plus1ex \@minus.2ex}%
    {-1em}%
    {\normalfont\normalsize\bfseries}}
\makeatother

\usepackage[compact]{titlesec}

\AtBeginDocument{%
    \addtolength\abovedisplayskip{-5pt}
    \addtolength\belowdisplayskip{-5pt}
}

\makeatother

  \providecommand{\definitionname}{Definition}
  \providecommand{\remarkname}{Remark}
\providecommand{\theoremname}{Theorem}

\begin{document}

\title{\textbf{Exact and Efficient Inference for Partial Bayes Problems}}

\author{Yixuan Qiu\\
{\normalsize{}Department of Statistics, Purdue University, }\texttt{\small{}yixuanq@purdue.edu}
\and Lingsong Zhang\\
{\normalsize{}Department of Statistics, Purdue University, }\texttt{\small{}lingsong@purdue.edu}
\and Chuanhai Liu\\
{\normalsize{}Department of Statistics, Purdue University, }\texttt{\small{}chuanhai@purdue.edu}}
\maketitle
\begin{abstract}
Bayesian methods are useful for statistical inference. However, real-world
problems can be challenging using Bayesian methods when the data analyst
has only limited prior knowledge. In this paper we consider a class
of problems, called Partial Bayes problems, in which the prior information
is only partially available. Taking the recently proposed Inferential
Model approach, we develop a general inference framework for Partial
Bayes problems, and derive both exact and efficient solutions. In
addition to the theoretical investigation, numerical results and real
applications are used to demonstrate the superior performance of the
proposed method.\bigskip{}

\newpage{}
\end{abstract}

\section{Introduction}

\label{sec:introduction}

In many real-world statistical problems, the information that is available
to the data analysts can be organized in a hierarchical structure.
That is, there exists some past experience about the parameter(s)
of interest, and data relevant to the parameter(s) are also collected.
For this type of problems, the standard approach to statistical inference
is the Bayesian framework. However, in many applications, the data
analysts have only limited prior knowledge. For instance, the prior
information may be insufficient to form a known distribution, so that
data analysts need to assume some unknown distributional components
in the Bayesian setting. This class of problems has brought many challenges
to statisticians; see for example \citet{lambert1986single,meaux2002statistical,moreno2003bayesian}.
To systematically study such problems that involve partial prior information,
in this article we refer to them as \emph{Partial Bayes problems},
in order to highlight their nature that there exists only partial
information in the Bayesian prior distribution.

Partial Bayes problems have drawn a lot of attention in statistics
literature. One popular type of Partial Bayes problems refers to the
case where there exists an unknown prior distribution, either parametric
or non-parametric, in a Bayesian hierarchical model. A very popular
approach to this type of models is known as the Empirical Bayes, which
has been first proposed by \citet{robbins1956} for handling the case
with non-parametric prior distributions, and later by \citet{efron1971limiting,efron1972empirical,efronmorris1972,efron1973stein,efron1975data}
for parametric prior distributions. Another kind of Partial Bayes
problems was studied by \citet{xie2013incorporating}, in which the
joint prior distribution of a parameter vector is missing, but some
marginal distributions are known. For clarity, we will refer to this
type as the marginal prior problem. In \citet{xie2013incorporating},
the solution to the marginal prior problem is based on the Confidence
Distribution approach \citep{xie2012confidence}, which provides a
unified framework for meta-analysis.

The Empirical Bayes and Confidence Distribution approaches both have
successful real-world applications. However, one fundamental problem
in scientific research, the exact inference about the parameter of
interest, remains to be an open question for Partial Bayes problems.
As pointed out by many authors \citep{morris1983parametric,laird1987empirical,carlin1990approaches},
Empirical Bayes in general underestimates the associated uncertainty
of the interval estimators, so these authors have proposed various
methods to correct the bias of the coverage rate. However, even if
they have shown better performance, the target coverage rates are
still approximately achieved for such methods. The same issue happens
in the Confidence Distribution framework. Confidence Distribution
provides a novel way to combine different inference results, but these
individual inferences may or may not be exact. All of these indicate
that the exact inference for Partial Bayes problems is highly non-trivial.

Recently, the Inferential Model \citep{martin2013inferential,martin2015conditional,martin2014marginal}
is proposed as a new framework for statistical inference, which not
only provides Bayesian-like probabilistic measures of uncertainty
about the parameter, but also has an automatic long-run frequency
calibration property. In this paper, we use this framework to derive
interval estimators for the parameters of interest in Partial Bayes
problems, and demonstrate their important statistical properties including
the exactness and efficiency. When compared with other approaches,
we refer to the proposed estimators as Partial Bayes solutions for
brevity.

The remaining part of this article is organized as follows. In Section
\ref{sec:motivating_example} we study a hierarchical normal-means
model as a motivating example of Partial Bayes problems. In Section
\ref{sec:review_im} we provide a brief review of the Inferential
Model framework as the theoretical foundation of our analysis. Section
\ref{sec:partial_bayes} is the main part of this article, where we
introduce a general framework for studying Partial Bayes problems,
and deliver our major theoretical results. We revisit some popular
Partial Bayes models in Section \ref{sec:popular_models}, are conduct
simulation studies in Section \ref{sec:simulation} to numerically
compare the proposed solutions with other methods. In Section \ref{sec:application}
we consider an application to a basketball game dataset, and finally
in Section \ref{sec:conclusion} we conclude with a few remarks. Proofs
of theoretical results are given in the appendix.

\section{A Motivating Example}

\label{sec:motivating_example}

In this section, we use a motivating example to demonstrate what a
typical Partial Bayes problem is, and how its solution differs from
the existing method. Consider the well-known normal hierarchical model
for the observed data $X=(X_{1},\ldots,X_{n})'$. The model introduces
$n$ unobservable means $\mu_{1},\ldots,\mu_{n}$, one for each observation,
and assumes that conditional on $\mu_{i}$'s, $X_{i}$'s are mutually
independent with $X_{i}|\{\mu_{1},\ldots,\mu_{n}\}\sim\mathsf{N}(\mu_{i},\sigma^{2})$
for $i=1,\ldots,n$, where the common variance $\sigma^{2}$ is known.
In addition, all the $\mu_{i}$'s are i.i.d. with $\mu_{i}\sim\mathsf{N}(\mu,\tau^{2})$
for $i=1,\ldots,n$, where the variance $\tau^{2}$ is known but the
mean $\mu$ is an unknown hyper-parameter.

The problem of interest here is to make inference about the individual
means $\mu_{i}$, and for simplicity we focus on $\mu_{1}$ without
loss of generality. The aim of inference is to construct an interval
estimator for $\mu_{1}$ that satisfies the following conditions:
using the terminology in \citet{morris1983parametric}, a sample-based
interval $C_{\alpha}(X)$ is an interval estimator for $\mu_{1}$
with $100(1-\alpha)\%$ confidence level, if it satisfies $P_{\mu_{1},X}(C_{\alpha}(X)\ni\mu_{1})\ge1-\alpha$
for all $\mu$, where the probability that indicates the coverage
rate is computed over the joint distribution of $(X,\mu_{1})$.

The standard Empirical Bayes approach to this problem can be found
in \citet{efron2010large}. It computes the MLE of $\mu$, $\hat{\mu}=\overline{X}$,
from the observed data. Plugging $\hat{\mu}$ back into the prior
in place of $\mu$, Empirical Bayes proceeds with the standard Bayesian
procedure to provide an approximate posterior distribution of $\mu_{1}$,
$\mu_{1}|X\overset{\cdot}{\sim}\mathsf{N}\left((1-\omega)X_{1}+\omega\overline{X},(1-\omega)\sigma^{2}\right)$,
where $\omega=\sigma^{2}/(\tau^{2}+\sigma^{2})$, and the notation
``$\overset{\cdot}{\sim}$'' indicates that the distribution is
approximate. Accordingly, the $100(1-\alpha)\%$ Empirical Bayes interval
estimator for $\mu_{1}$ is obtained as
\[
(1-\omega)X_{1}+\omega\overline{X}\pm z_{\alpha/2}\sigma\sqrt{1-\omega},
\]
where $z_{\alpha/2}$ is the $1-\alpha/2$ quantile of the standard
normal distribution.

\begin{figure}[h]
\begin{centering}
\includegraphics[width=0.7\textwidth]{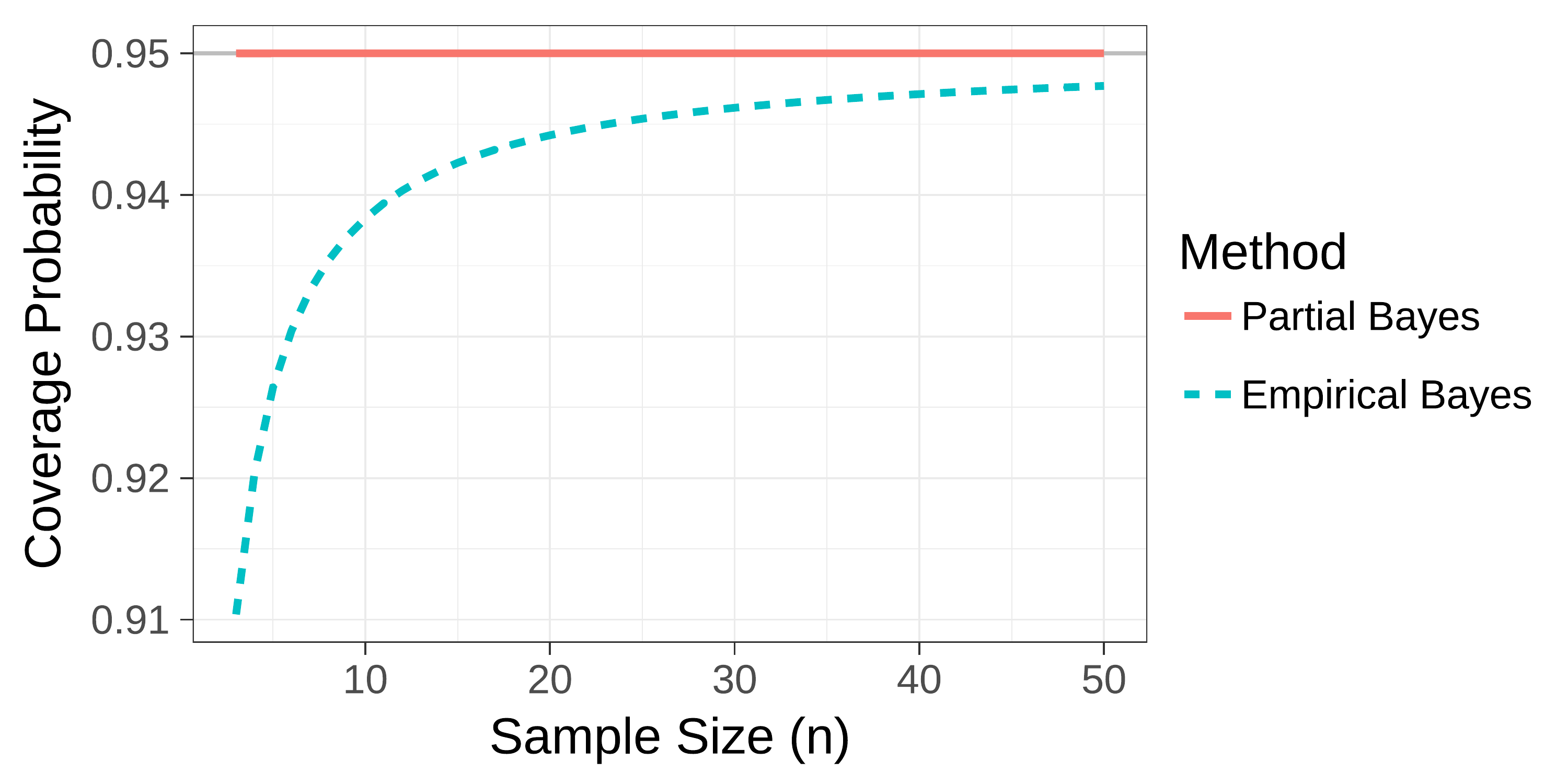}
\par\end{centering}
\caption{\setlength{\belowcaptionskip}{-10pt} The coverage probabilities of
Empirical Bayes (blue dashed curve) and Partial Bayes (red solid line)
as a function of $n$. The line for Partial Bayes is exactly positioned
at the 0.95 level, indicating that it achieves the nominal coverage
rate exactly for all $n$. \label{fig:coverage_rate}}
\end{figure}

The Partial Bayes solution, derived in Section \ref{subsec:normal_example_1},
has a slightly different formula:
\begin{equation}
(1-\omega)X_{1}+\omega\overline{X}\pm z_{\alpha/2}\sigma\sqrt{1-\omega(n-1)/n}.\label{eq:motivating_example_interval}
\end{equation}
Compared with Empirical Bayes, the proposed interval has the same
center but is slightly wider for small $n$. For a numerical illustration,
we fix $\alpha$ to be 0.05, and take $\sigma^{2}=\tau^{2}=1$. Figure
\ref{fig:coverage_rate} shows the theoretical coverage rates of both
the Empirical Bayes solution and the Partial Bayes solution as a function
of $n$. It can be seen that the coverage probability of the Empirical
Bayes interval is less than the nominal value $1-\alpha$, and is
close to the target only when $n$ is sufficiently large. On the contrary,
the Partial Bayes solution correctly matches the nominal coverage
rate for all $n$.

\section{A Brief Review of Inferential Models}

\label{sec:review_im}

Since our inference for Partial Bayes problems is based on the recently
developed Inferential Models, in this section we provide a brief introduction
to this new framework, with more details given in \citet{martin2013inferential}.
Inferential Model is a new framework designed for exact and efficient
statistical inference. The exactness of Inferential Models guarantees
that under a particular definition, the inference made by Inferential
Models has a controlled probability of error, for example, in hypothesis
testing problems the Type I error should be no greater than a pre-specified
level. In addition, Inferential Models provide a systematic way to
combine information in the data for efficient statistical inference.

Formally, Inferential Models draw statistical conclusions on an assertion
$A$, a subset of the parameter space, about the parameter of interest
$\theta$. For example, the subset $A=\{0\}$ stands for the assertion
$\theta=0$, and $A=(1,+\infty)$ corresponds to $\theta>1$. In the
Inferential Model framework, two quantities are used to represent
the knowledge about $A$ contained in the data: the \emph{belief function},
which describes how much evidence in the data supports the claim that
``$A$ is true'', and the \emph{plausibility function}, which quantifies
how much evidence does not support the claim that ``$A$ is false''.

Like Fisher's fiducial inference, Inferential Models make use of auxiliary
or unobserved random variables to represent the sampling model. In
order to have meaningful probabilistic inferential results, unlike
Fisher's fiducial inference, Inferential Models predict unobserved
realizations of the auxiliary variables using random sets, and propagate
such uncertainty to the space of $\theta$. Technically, Inferential
Model is formulated as a three-step procedure to produce the inferential
results:

\paragraph{Association step}

This step specifies an association function $X=a(\theta,U)$ to connect
the parameter $\theta\in\Theta$, the observed data $X\in\mathbb{X}$,
and the unobserved auxiliary random variable $U\in\mathbb{U}$ with
$U$ following a known distribution $\mathsf{P}_{U}$. This relationship
implies that the randomness in the data is represented by an auxiliary
variable $U$.

\paragraph{Prediction step}

Let $u^{*}$ be the true but unobserved value of $U$ that ``generates''
the data. This step constructs a \emph{valid} predictive random set,
$\mathcal{S}$, to predict $u^{*}$. $\mathcal{S}$ is valid if the
quantity $Q_{\mathcal{S}}(u^{*})=P_{\mathcal{S}}(\mathcal{S}\ni u^{*})$,
interpreted as the probability that $\mathcal{S}$ successfully covers
$u^{*}$, satisfies the condition $P_{U}(Q_{\mathcal{S}}(U)\ge1-\alpha)\ge1-\alpha$,
where $U\sim\mathsf{P}_{U}$.

\paragraph{Combination step}

This step transforms the uncertainty from the $\mathbb{U}$ space
to the $\Theta$ space by defining $\Theta_{x}(\mathcal{S})=\bigcup_{u\in\mathcal{S}}\Theta_{x}(u)=\bigcup_{u\in\mathcal{S}}\{\theta:x=a(u,\theta)\}$,
a mapping from $U$ back to $\theta$ after incorporating the uncertainty
represented by $\mathcal{S}$. Then for an assertion $A$, its belief
function is defined as $\mathsf{bel}_{x}(A)=P\{\Theta_{x}(\mathcal{S})\subseteq A|\Theta_{x}(\mathcal{S})\ne\varnothing\}$,
and similarly, its plausibility function is defined as $\mathsf{pl}_{x}(A)=1-\mathsf{bel}_{x}(A^{c})$.\bigskip{}

The plausibility function is very useful to derive frequentist-like
confidence regions for the parameter of interest \citep{martin2015plausibility}.
If we let $A$ be a singleton assertion $A=\{\theta\}$ and denote
$\mathsf{pl}_{x}(\theta)\equiv\mathsf{pl}_{x}(\{\theta\})$, then
a $100(1-\alpha)\%$ frequentist-like confidence region, which is
termed as \emph{plausibility region} in Inferential Model (or \emph{plausibility
interval} as a special case), is given by $\mathsf{PR}_{x}(\alpha)=\{\theta:\mathsf{pl}_{x}(\theta)>\alpha\}$.
In Inferential Model, the exactness of the inference is formally termed
as \emph{validity}. For example, the validity property of Inferential
Model guarantees that the above region $\mathsf{PR}_{x}(\alpha)$
has at least $100(1-\alpha)\%$ long-run coverage probability.

It is worth mentioning that Inferential Models also have a number
of extensions for efficient inference. When the model has multiple
parameters but only some of them are of interest, the Marginal Inferential
Models (MIM, \citealp{martin2014marginal}) appropriately integrate
out the nuisance parameters. For models where the dimension of auxiliary
variables is higher than that of the parameters, the Conditional Inferential
Models (CIM, \citealp{martin2015conditional}) could be used to combine
information in the data such that efficient inference can be achieved.
Both MIM and CIM are used extensively in our development of exact
and efficient inference for Partial Bayes problems.

\section{Inference for Partial Bayes Problems}

\label{sec:partial_bayes}

In this section we build a general model framework for studying Partial
Bayes problems. The derivation of our interval estimator is described
in detail using the Inferential Model framework, and some of its key
statistical properties are also studied.

\subsection{Model Specification}

\label{subsec:model_specification}

Our attempt here is to provide a simple model framework that is general
enough to describe a broad range of Partial Bayes problems introduced
in Section \ref{sec:introduction}.

Let $X$ be the observed data, whose distribution $f$ relies on an
unknown parameter vector $\theta$. The information on $\theta$ that
comes from the collected data is expressed by the conditional distribution
of $X$ given the parameter: $X|\theta\sim f(x|\theta)$. In many
cases, we have prior knowledge about $\theta$ that can be characterized
as a prior distribution $\pi_{0}(\theta)$. When $\pi_{0}(\theta)$
is fully specified, standard Bayesian method can be used to derive
the posterior distribution of $\theta$. In other cases, there is
only partial prior information available. Formally, assume that the
parameter $\theta$ can be partitioned into two blocks, $\theta=(\tilde{\theta},\theta^{*})$,
so that the desirable fully-specified prior of $\theta$ can be accordingly
decomposed as $\pi_{0}(\theta)=\pi(\tilde{\theta}|\theta^{*})\pi^{*}(\theta^{*})$,
where $\pi(\tilde{\theta}|\theta^{*})$ is the conditional density
function of $\tilde{\theta}$ given $\theta^{*}$, and $\pi^{*}(\theta^{*})$
is the marginal distribution of $\theta^{*}$. We call the prior information
partial if only the conditional distribution $\tilde{\theta}|\theta^{*}\sim\pi(\tilde{\theta}|\theta^{*})$
is available, but $\pi^{*}(\theta^{*})$ is missing. In general, inference
is made on $\tilde{\theta}$ or a component of $\tilde{\theta}$,
\emph{i.e.}, $\tilde{\theta}$ can be further partitioned into $\tilde{\theta}=(\eta,\xi)$,
with $\eta$ denoting the parameter of interest and $\xi$ denoting
the additional nuisance parameters. In this article we focus on the
case that $\eta$ is a scalar, which is of interest for many practical
problems. For better presentation, we summarize these concepts and
the proposed model structure in the following table:

\begin{center}
\begin{tabular}{ll}
\toprule 
\addlinespace[0.1em]
Sampling model & $X|\theta\sim f(x|\theta)$\tabularnewline\addlinespace[0.1em]
\addlinespace[0.1em]
Parameter partition & $\theta=(\tilde{\theta},\theta^{*})$, $\tilde{\theta}=(\eta,\xi)$\tabularnewline\addlinespace[0.1em]
\addlinespace[0.1em]
Partial prior & $\tilde{\theta}|\theta^{*}\sim\pi(\tilde{\theta}|\theta^{*})$\tabularnewline\addlinespace[0.1em]
\addlinespace[0.1em]
Component without prior & $\theta^{*}$\tabularnewline\addlinespace[0.1em]
\addlinespace[0.1em]
Parameter of interest & $\eta$\tabularnewline\addlinespace[0.1em]
\bottomrule
\end{tabular}
\par\end{center}

Despite its simplicity, the above model includes the well-known hierarchical
models as an important class of practically useful models. Moreover,
the formulation goes beyond the hierarchical models, and also includes
the marginal prior problem. As described in Section \ref{sec:introduction},
our target of inference is to construct a sample-based interval $C(X)$
that satisfies some validity conditions. Specifically, the following
two types of validity properties are considered:
\begin{defn}
\label{def:unconditional_validity}$C(X)$ is said to be an \emph{unconditionally
valid} interval estimator for $\eta$ with $100(1-\alpha)\%$ confidence
level, if $P_{X,\theta}(C(X)\ni\eta)\ge1-\alpha$ for all $\pi^{*}(\theta^{*})$,
where the probability is computed over the joint distribution of $(X,\theta)$.
\end{defn}
\begin{defn}
\label{def:conditional_validity}$C(X)$ is said to be a \emph{conditionally
valid} interval estimator for $\eta$ given $H(X)$ with $100(1-\alpha)\%$
confidence level, if $P_{X,\theta|H(X)}(C(X)\ni\eta|H(X)=h)\ge1-\alpha$
for all $\pi^{*}(\theta^{*})$ and $h$, where $H(X)$ is a statistic
of the data, and the probability is computed over the joint distribution
of $(X,\theta)$ given $H(X)=h$.
\end{defn}
Definition \ref{def:unconditional_validity} is a rephrasing of the
validity condition in \citet{morris1983parametric}, and Definition
\ref{def:conditional_validity} comes from \citet{carlin1990approaches}.
It should be noted that the second condition is stronger than the
first, since it can be reduced to Definition \ref{def:unconditional_validity}
by averaging over $H(X)$. In this article, we aim to produce the
second type of interval estimators, but the first validity property
is studied when different interval estimators for $\eta$ are compared
with each other.

\subsection{Inferential Models for Partial Bayes Problems}

\label{subsec:IM_for_PB}

In this section we describe a procedure to analyze Partial Bayes problems
in the Inferential Model framework, and develop intermediate results
that are used to derive the proposed interval estimator in Section
\ref{subsec:validity}. The procedure consists of the three steps
introduced in Section \ref{sec:review_im}, and outputs a plausibility
function for $\eta$, the parameter of interest.

\subsubsection{The Association Step}

\label{subsec:A_step}

The association step has three sub-steps, and we highlight their tasks
at the beginning of each sub-step.

\paragraph{Constructing data and prior associations}

The first association equation comes from the data sampling model
$X|\theta\sim f(x|\theta)$, for which we write $X=a_{1}(\theta,W_{1})$,
where $a_{1}(\cdot)$ is the ``data association'' function, and
$W_{1}$ is an unobservable auxiliary variable that has a known distribution.
Since $\theta$ can be partitioned into $\theta=(\tilde{\theta},\theta^{*})$
with $\tilde{\theta}|\theta^{*}\sim\pi(\tilde{\theta}|\theta^{*})$,
the equation that represents this partial information can be written
as $\tilde{\theta}=a_{2}(\theta^{*},W_{2})$, where $a_{2}(\cdot)$
is the ``prior association'' function, and $W_{2}$ is another auxiliary
variable independent of $W_{1}$. Substituting the prior association
into the data association, we get $X=a_{1}((a_{2}(\theta^{*},W_{2}),\theta^{*}),W_{1})$.
To avoid the over-complicated notations, we simply write this relation
as $X=a(\theta^{*},W)$, where $W=(W_{1},W_{2})$.

As described in Section \ref{subsec:model_specification}, we are
only interested in an element of the $\tilde{\theta}$ vector, so
we assume that $\tilde{\theta}=a_{2}(\theta^{*},W_{2})$ can be equivalently
decomposed as $\eta=a_{\eta}(\theta^{*},V_{\eta})$ and $\xi=a_{\xi}(\theta^{*},V_{\xi})$,
where $a_{\eta}(\cdot)$ and $a_{\xi}(\cdot)$ are the decomposed
associations and $W_{2}=(V_{\eta},V_{\xi})$. Therefore, the model
for Partial Bayes problems can be summarized by the following system
of three equations:
\begin{equation}
X=a(\theta^{*},W),\ \eta=a_{\eta}(\theta^{*},V_{\eta}),\ \text{and}\ \xi=a_{\xi}(\theta^{*},V_{\xi}).\label{eq:PB_association}
\end{equation}
Note that $\xi$ can be regarded as a nuisance parameter, and \eqref{eq:PB_association}
is ``regular'' in the sense of Definition 2 of \citet{martin2014marginal}.
Then according to the general theory of MIM in that paper (Theorems
2 and 3), the third equation in \eqref{eq:PB_association} can be
ignored without loss of efficiency.

\paragraph{Decomposing data association}

Next, since the sample $X$ usually contains multiple observations,
the dimension of $W$ can often be very high. In order to reduce the
number of auxiliary variables, assume that the relationship $X=a(\theta^{*},W)$
admits a decomposition
\begin{equation}
T(X)=a_{T}(\theta^{*},\tau(W)),\ \text{and}\ H(X)=\rho(W)\label{eq:CIM_decomposition}
\end{equation}
for one-to-one mappings $x\mapsto(T(x),H(x))$ and $w\mapsto(\tau(w),\rho(w))$.
\citet{martin2015conditional} shows that this decomposition broadly
exists for a large number of models, and in case that \eqref{eq:CIM_decomposition}
is not available, we simply write $H(X)=1$ and $\rho(W)=1$. The
equation \eqref{eq:CIM_decomposition} implies that when the collected
data have a realization $x$, the auxiliary variable $W_{H}\coloneqq\rho(W)$
is fully observed with the value $h\coloneqq H(x)$. By conditioning
on $W_{H}=h$, we obtain the following two conditional associations
\begin{align}
T(X)=a_{T}(\theta^{*},W_{T}), & \quad W_{T}\coloneqq\tau(W)\sim\mathsf{P}_{W_{T}|h},\label{eq:cond_association_T}\\
\eta=a_{\eta}(\theta^{*},V_{\eta}), & \quad V_{\eta}\sim\mathsf{P}_{V_{\eta}|h},\label{eq:cond_association_eta}
\end{align}
where the notation $Z\sim\mathsf{P}_{Z|h}$ means that the random
variable $Z$ has a distribution $\mathsf{P}_{Z|h}$ given $W_{H}=h$.
In the rest of Section \ref{subsec:IM_for_PB}, when we discuss the
distribution of a random variable that depends on $W_{T}$ or $V_{\eta}$,
the condition $W_{H}=h$ is implicitly added.

\paragraph{Obtaining the final association}

Finally, to make inference about $\eta$, the unknown quantity $\theta^{*}$
needs to be marginalized out of the equations. We seek a real-valued
continuous function $b(\cdot,\cdot)$ such that when its first argument
is fixed to some value $t$, the mapping $\eta\mapsto b(t,\eta)$
is one-to-one. At the current stage we simply take $b$ as an arbitrary
function, and we defer the discussion of its optimal choice in Section
\ref{subsec:efficiency}. As a result, associations \eqref{eq:cond_association_T}
and \eqref{eq:cond_association_eta} are equivalent to
\begin{align}
T(X)=a_{T}(\theta^{*},W_{T}),\label{eq:cond_association_T_2}\\
b(T(X),\eta)=W_{b}(\theta^{*}), & \quad W_{b}(\theta^{*})\coloneqq b(a_{T}(\theta^{*},W_{T}),a_{\eta}(\theta^{*},V_{\eta})).\label{eq:final_association}
\end{align}
Conditional on $\theta^{*}$, $W_{b}(\theta^{*})$ is a random variable
whose c.d.f. $F_{W_{b}(\theta^{*})|h}$ is indexed by the unknown
parameter $\theta^{*}$. If the function $b$ is chosen such that
$\theta^{*}$ has only little effect on $F_{W_{b}(\theta^{*})|h}$,
the first equation \eqref{eq:cond_association_T_2} provides little
or even no information about $\eta$, and hence it can be ignored
according to the theory of MIM. The final association equation \eqref{eq:final_association}
thus completes the association step.

\subsubsection{The Prediction Step}

\label{subsec:P_step}

The aim of the this step is to introduce a predictive random set $\mathcal{S}_{h}$
conditional on $W_{H}=h$ that can predict $W_{b}(\theta^{*})$ with
high probability. The following two situations are considered.

The first situation is that $W_{b}(\theta^{*})$ is in fact free of
$\theta^{*}$. This can be easily achieved if $\theta^{*}$ has the
same dimension as $\eta$, and if the mapping $\eta=a_{\eta}(\theta^{*},V_{\eta})$
can be inverted as $\theta^{*}=a_{\theta^{*}}(\eta,V_{\eta})$. To
verify this, plug $\theta^{*}=a_{\theta^{*}}(\eta,V_{\eta})$ into
\eqref{eq:cond_association_T}, and we obtain $T(X)=a_{T}(a_{\theta^{*}}(\eta,V_{\eta}),W_{T})$,
which reduces to a univariate Inferential Model problem that has a
well-defined solution.

The second situation is more general and thus more challenging, in
which case $F_{W_{b}(\theta^{*})|h}$ relies on the unknown parameter
$\theta^{*}$. Typically this occurs when the dimension of $\theta^{*}$
is higher than that of $\eta$. To deal with this issue, we generalize
the Definition 5 of \citet{martin2014marginal} to define the concept
of stochastic bounds for tails.
\begin{defn}
\label{def:tail_bounds} Let $Z$ and $Z^{*}$ be two random variables
with c.d.f. $F_{Z}$ and $F_{Z^{*}}$ respectively, and denote by
$\mathrm{med}(Z)$ the median of $Z$. $Z$ is said to be stochastically
bounded by $Z^{*}$ in tails if $F_{Z}(z)\le F_{Z^{*}}(z)$ for $z<\mathrm{med}(Z)$,
and $F_{Z}(z)\ge F_{Z^{*}}(z)$ for $z>\mathrm{med}(Z)$.
\end{defn}
\noindent The difference between this definition and the one in the
literature is that here the medians of $Z$ and $Z^{*}$ are not required
to be zero.

Assume that we have found a random variable $W_{b}^{*}$ such that
given $W_{H}=h$, $W_{b}(\theta^{*})$ is stochastically bounded by
$W_{b}^{*}$ in tails for any $\theta^{*}$. Note that the first situation
discussed earlier can be viewed as a special case, since any random
variable is stochastically bounded by itself in tails. To shorten
the argument, we only consider this more general case for later discussion.
There are various ways to construct such a random variable $W_{b}^{*}$,
see the examples in \citet{martin2014marginal}. Here we provide a
simple approach, by defining the c.d.f. to be
\[
F_{W_{b}^{*}|h}(z)=\left\{ \begin{array}{ll}
\sup_{\theta^{*}}F_{W_{b}(\theta^{*})|h}(z), & z<m_{h}\\
\frac{1}{2}, & z=m_{h}\\
\inf_{\theta^{*}}F_{W_{b}(\theta^{*})|h}(z), & z>m_{h}
\end{array}\right.,\ {\textstyle m_{h}=F_{W_{b}(\theta^{*})|h}^{-1}\left(\frac{1}{2}\right)},
\]
provided that the resulting function is a c.d.f..

Given $F_{W_{b}^{*}|h}$, a standard conditional predictive random
set $\mathcal{S}_{h}$ can be chosen for the prediction of $W_{b}(\theta^{*})$.
For the purpose of constructing two-sided interval estimators, we
first define the generalized c.d.f. of a random variable $Z$ as $F_{Z}^{-1}(u)=\inf\{x:F_{Z}(x)\ge u\}$,
and then construct $\mathcal{S}_{h}$ as follows:
\begin{equation}
\mathcal{S}_{h}=\left\{ F_{W_{b}^{*}|h}^{-1}(u'):\vert u'-0.5\vert<\vert U_{\mathcal{S}}-0.5\vert,u'\in(0,1)\right\} ,\ U_{\mathcal{S}}\sim\mathsf{Unif}(0,1).\label{eq:random_set}
\end{equation}
This completes the prediction step, and other choices of the predictive
random set for different purposes are discussed in \citet{martin2013inferential}.

\subsubsection{The Combination Step}

\label{subsec:C_step}

In what follows, to avoid notational confusions we use $\eta$ to
represent the parameter of interest as a random variable, and denote
by $\tilde{\eta}$ the possible values of $\eta$. In the final combination
step, denote by $\Theta_{T(x)}(w)$ the set of $\tilde{\eta}$ values
that satisfy the association equation \eqref{eq:final_association}
with $T(X)=T(x)$ and $W_{b}(\theta^{*})=w$, \emph{i.e.}, $\Theta_{T(x)}(w)=\{\tilde{\eta}:b(T(x),\tilde{\eta})=w\}$,
and define $\Theta_{T(x)}(\mathcal{S}_{h})=\bigcup_{s\in\mathcal{S}_{h}}\Theta_{T(x)}(s)$.
Then the conditional plausibility function for $\eta$ is obtained
as
\begin{equation}
\mathsf{cpl}_{T(x)|h}(\tilde{\eta})=1-P_{\mathcal{S}_{h}}\left(\Theta_{T(x)}(\mathcal{S}_{h})\subseteq(-\infty,\tilde{\eta})\cup(\tilde{\eta},+\infty)\right),\label{eq:conditional_plausibility}
\end{equation}
which completes the combination step.

\subsection{Interval Estimator and Validity of Inference}

\label{subsec:validity}

In Section \ref{subsec:C_step} a conditional plausibility function
for the $\eta$ parameter has been derived under the Inferential Model
framework, and in this section it is used to construct the proposed
interval estimator. Similar to the construction of plausibility region
introduced in Section \ref{sec:review_im}, we define the following
set-valued function of $x$:
\begin{equation}
C_{\alpha}(x)=\{\tilde{\eta}:\text{\ensuremath{\mathsf{cpl}}}_{T(x)|h}(\tilde{\eta})\ge\alpha\}.\label{eq:interval_estimator}
\end{equation}
From \eqref{eq:conditional_plausibility} it can be seen that $\mathsf{cpl}_{T(x)|h}(\tilde{\eta})$
depends on the data on two aspects: the random set $\mathcal{S}_{h}$
depends on $h=H(x)$, and the association function $\Theta_{T(x)}(w)$
depends on $T(x)$. As a result, we define our Partial Bayes interval
estimator for $\eta$ to be $C_{\alpha}(X)$, obtained by plugging
the random sample $X$ into $C_{\alpha}(x)$.

In the typical case that $\eta$ is a fixed value, the Inferential
Model theory guarantees that $C_{\alpha}(X)$ is a valid $100(1-\alpha)\%$
frequentist confidence interval for $\eta$. However in our case,
the joint distribution of the parameter and data is considered, as
in Definitions \ref{def:unconditional_validity} and \ref{def:conditional_validity}.
Therefore, the validity of $C_{\alpha}(X)$ does not automatically
follow from the Inferential Model theory, and hence needs to be studied
separately. The result is summarized as Theorem \ref{thm:cond_validity}.
\begin{thm}
\label{thm:cond_validity}With $H(X)$ defined in \eqref{eq:CIM_decomposition},
$C_{\alpha}(X)$ is a conditionally valid interval estimator for $\eta$
given $H(X)$ with $100(1-\alpha)\%$ confidence level.
\end{thm}
Recall that if the decomposition \eqref{eq:CIM_decomposition} is
unavailable, we will take $H(X)=1$ and $\rho(W)=1$. In such cases,
Theorem \ref{thm:cond_validity} reduces to the unconditional result
corresponding to Definition \ref{def:unconditional_validity}.

\subsection{Optimality and Efficiency}

\label{subsec:efficiency}

Theorem \ref{thm:cond_validity} states that the proposed interval
estimator $C_{\alpha}(X)$ defined in \eqref{eq:interval_estimator}
satisfies the validity condition. Another important property, the
efficiency of the estimator, is discussed in this section. We claim
two facts about the proposed interval estimator:
\begin{enumerate}
\item If $\pi^{*}(\theta^{*})$ is known, then with a slight modification
to the predictive random set $\mathcal{S}_{h}$, the optimal interval
estimator $C_{\alpha}^{o}(X)$ can be constructed.
\item If $\pi^{*}(\theta^{*})$ is unknown, then under some mild conditions,
$C_{\alpha}(X)$ can approximate $C_{\alpha}^{o}(X)$ well. The discussion
also guides the choice of the $b$ function in \eqref{eq:final_association}.
\end{enumerate}
First consider the ideal scenario that $\pi^{*}(\theta^{*})$, the
marginal distribution of $\theta^{*}$, is known, in which case a
full prior distribution for $\theta$ is available. On one hand, it
is well known that given a fully-specified prior distribution, the
optimal inference for the parameter is via its posterior distribution
given the data. On the other hand, given this new information, the
approach introduced in Section \ref{subsec:IM_for_PB} can still be
used to derive an interval estimator, with some slight modifications
shown below. Later this result is compared with the Bayesian solution.

Let $\theta^{*}=U,\ U\sim\pi^{*}(\theta^{*})$ be the association
equation for the marginal distribution of $\theta^{*}$. Combining
it with \eqref{eq:cond_association_T_2} and \eqref{eq:final_association},
we obtain the following three associations:
\begin{equation}
\theta^{*}=U,\ T(X)=Z_{T},\ \text{and}\ b(T(X),\eta)=W_{b},\label{eq:association_optimal}
\end{equation}
where $Z_{T}=a_{T}(U,W_{T})$ and $W_{b}=b(Z_{T},a_{\eta}(U,V_{\eta}))$.
Again, the second equation implies that given the data $x$, $Z_{T}$
is fully observed with value $t\coloneqq T(x)$, so the auxiliary
variable $W_{b}$ can be predicted using its conditional distribution
given $W_{H}=h$ and $Z_{T}=t$, which we denote by $F_{W_{b}|h,t}$.
Similar to the prediction step in Section \ref{subsec:P_step}, we
construct a predictive random set $\mathcal{S}_{h,t}$ for $W_{b}$
by replacing $F_{W_{b}^{*}|h}^{-1}$ with $F_{W_{b}|h,t}^{-1}$ in
formula \eqref{eq:random_set}, and proceed with the same combination
step to obtain
\[
\mathsf{cpl}_{T(x)|h,t}(\tilde{\eta})=1-P_{\mathcal{S}_{h,t}}\left(\Theta_{T(x)}(\mathcal{S}_{h,t})\subseteq(-\infty,\tilde{\eta})\cup(\tilde{\eta},+\infty)\right).
\]
As a result, the interval estimator for $\eta$ is obtained as $C_{\alpha}^{o}(X)$,
where $C_{\alpha}^{o}(x)=\{\tilde{\eta}:\text{\ensuremath{\mathsf{cpl}}}_{T(x)|h,t}(\tilde{\eta})\ge\alpha\}$.
Comparing the $\mathsf{cpl}_{T(x)|h,t}(\tilde{\eta})$ function that
defines $C_{\alpha}^{o}(X)$ and the $\mathsf{cpl}_{T(x)|t}(\tilde{\eta})$
function in \eqref{eq:conditional_plausibility}, it can be seen that
they only differ in the distributions assigned to the predictive random
sets. The following theorem shows that with this slight change, $C_{\alpha}^{o}(X)$
matches the Bayesian posterior credible interval.
\begin{thm}
\label{thm:optimality}Assuming that $\pi^{*}(\theta^{*})$ is known
and $\eta$ has a continuous distribution function $F_{\eta|x}$ given
$X=x$, then $C_{\alpha}^{o}(X)$ is optimal in the sense that it
matches the Bayesian posterior credible interval, i.e., $C_{\alpha}^{o}(x)=\left(F_{\eta|x}^{-1}(\alpha/2),F_{\eta|x}^{-1}(1-\alpha/2)\right)$.
\end{thm}
Theorem \ref{thm:optimality} implies that, by choosing a proper predictive
random set $\mathcal{S}_{h,t}$ for the $W_{b}$ auxiliary variable,
the inference result can attain the optimality. This fact implies
that even when $\pi^{*}(\theta^{*})$ is missing, as long as there
exists a predictive random set close to $\mathcal{S}_{h,t}$, the
resulting interval estimator would be as efficient as the optimal
one, at least approximately.

Recall that the optimal predictive random set $\mathcal{S}_{h,t}$
is induced by the distribution $F_{W_{b}|h,t}$, and when $\pi^{*}(\theta^{*})$
is missing, only $F_{W_{b}(\theta^{*})|h}$ is available. Therefore,
the next question is to find out the conditions under which $F_{W_{b}(\theta^{*})|h}$
is close to $F_{W_{b}|h,t}$. Since they are both conditional on $W_{H}=h$,
to simplify the analysis we remove this condition from both distributions,
and then study the closeness between $F_{W_{b}(\theta^{*})}$ and
$F_{W_{b}|t}$, where $F_{W_{b}(\theta^{*})}$ is the c.d.f. of $W_{b}(\theta^{*})$
defined in \eqref{eq:final_association}, and $F_{W_{b}|t}$ stands
for the distribution of $W_{b}$ defined in \eqref{eq:association_optimal}
given $Z_{T}=t$.

In most real applications, the association relation for $T(X)$ changes
with the data size $n$. To emphasize the dependence on $n$, in what
follows we write $W_{b_{n}}(\theta^{*})$, $Z_{T_{n}}$, and $W_{b_{n}}$
in place of $W_{b}(\theta^{*})$, $Z_{T}$, and $W_{b}$, respectively.
The following definition from \citet{XIONG20083249} is needed to
study the large sample property of a conditional distribution.
\begin{defn}
Given two sequences of random variables $X_{n}$ and $Y_{n}$, the
conditional distribution function of $X_{n}$ given $Y_{n}$, a random
c.d.f. denoted by $F_{X_{n}|Y_{n}}$, is said to converge weakly to
a non-random c.d.f. $F_{Z}$ in probability, denoted by $X_{n}|Y_{n}\overset{d.P}{\rightarrow}Z$,
if for every continuous point $z$ of $F_{Z}$, $F_{X_{n}|Y_{n}}(z)\overset{P}{\rightarrow}F_{Z}(z)$,
where $Z\sim F_{Z}$.

This definition is a generalization to the usual concept of weak convergence.
Then we have the following result:
\end{defn}
\begin{thm}
\label{thm:efficiency}Let $g_{n}$, $h_{n}$, and $p_{n}$ denote
the densities of $W_{b_{n}}$, $Z_{T_{n}}$, and $(W_{b_{n}},Z_{T_{n}})$,
respectively. Also define $l_{n}(w,z)=p_{n}(w,z)/[g_{n}(w)h_{n}(z)]$.
If (a) for fixed $u$, $a_{T}(u,W_{T_{n}})\overset{P}{\rightarrow}u$,
(b) $b(u,a_{\eta}(u,v))=v$, and (c) $l_{n}\rightarrow1$ pointwisely,
then $W_{b_{n}}|Z_{T_{n}}\overset{d.P}{\rightarrow}V_{\eta}$ and
$W_{b_{n}}(\theta^{*})\overset{d}{\rightarrow}V_{\eta}$, where $\theta^{*}$
in $W_{b_{n}}(\theta^{*})$ is seen as a fixed value.
\end{thm}
\begin{rem}
Conditions \emph{(a)} and \emph{(b)} are intentionally expressed in
a simple form. In fact they can be replaced by $a_{T}(u,W_{T_{n}})\overset{P}{\rightarrow}f_{1}(u)$
and $b(f_{1}(u),a_{\eta}(u,v))=f_{2}(v)$ where $f_{1}$ and $f_{2}$
are one-to-one functions, and the limiting distribution is changed
to $f_{2}(V_{\eta})$ accordingly.
\end{rem}
\begin{rem}
The three conditions are easy to check. Condition \emph{(a)} states
that $T(X)$ should be a consistent estimator for $\theta^{*}$ if
$\theta^{*}$ is seen as fixed. Condition \emph{(b)} guides the choice
of the $b$ function, \emph{e.g.} taking $b(t,\eta)=\inf\{v:a_{\eta}(t,v)=\eta\}$.
For condition \emph{(c)}, it is shown in the proof that $(W_{b_{n}},Z_{T_{n}})\overset{d}{\rightarrow}(V_{\eta},U)$,
and a sufficient condition for \emph{(c)} is that the density of $(W_{b_{n}},Z_{T_{n}})$
also converges to that of $(V_{\eta},U)$, which is satisfied by most
parametric models.

To summarize, Theorem \ref{thm:efficiency} indicates that $W_{b_{n}}(\theta^{*})$
and $W_{b_{n}}|Z_{T_{n}}$ converge to the same limiting distribution,
in which sense the random sets $\mathcal{S}_{h}$ and $\mathcal{S}_{h,t}$
have approximately identical distributions when $n$ is sufficiently
large. As a result, the proposed interval estimator $C_{\alpha}(X)$
defined in \eqref{eq:interval_estimator} can be seen as an approximation
to the optimal solution $C_{\alpha}^{o}(X)$. Combining Theorem \ref{thm:cond_validity}
and Theorem \ref{thm:efficiency}, it can be concluded that the proposed
interval estimator possesses the favorable properties of both validity
and efficiency.
\end{rem}

\section{Popular Models Viewed as Partial Bayes Problems}

\label{sec:popular_models}

In this section we apply the methodology in Section \ref{sec:partial_bayes}
to a collection of popular models viewed as Partial Bayes problems,
and show how their Partial Bayes solutions are developed.

\subsection{The Normal Hierarchical Model}

\label{subsec:normal_example}

The normal hierarchical model is extremely popular in the Empirical
Bayes literature, partly due to its simplicity and flexibility; see
for example \citet{efron1975data,morris1983parametric,casella1985introduction,efron2010large}.
The model setting has been given in Section \ref{sec:motivating_example},
and without loss of generality we set $\sigma^{2}=1$, since $X_{i}$'s
can always be scaled by a constant to achieve an arbitrary variance.
We will consider both the cases where $\tau^{2}$ is known and unknown,
and our parameter of interest is $\mu_{1}$. To summarize, we write

\begin{center}
\begin{tabular}{ll}
\toprule 
\addlinespace[0.1em]
Sampling model & $X|(\tilde{\theta},\theta^{*})\sim\prod_{i}\mathsf{N}(\mu_{i},\sigma^{2})$\tabularnewline\addlinespace[0.1em]
\addlinespace[0.1em]
Partial prior & $\tilde{\theta}=(\mu_{1},\mu_{2},\ldots,\mu_{n})$, $\tilde{\theta}|\theta^{*}\sim\prod_{i}\mathsf{N}(\mu,\tau^{2})$\tabularnewline\addlinespace[0.1em]
\addlinespace[0.1em]
Component without prior & $\theta^{*}=\left\{ \begin{array}{ll}
\mu, & \text{if }\tau\text{ is known}\\
(\mu,\tau^{2}), & \text{if }\tau\text{ is unknown}
\end{array}\right.$ \tabularnewline\addlinespace[0.1em]
\addlinespace[0.1em]
Parameter of interest & $\eta=\mu_{1}$\tabularnewline\addlinespace[0.1em]
\bottomrule
\end{tabular}
\par\end{center}

As a first step, this model can be expressed by the following association
equations: $\mu_{i}=\mu+\tau\varepsilon_{i}$ and $X_{i}=\mu_{i}+e_{i}$
for $i=1,\ldots,n$, where $\varepsilon_{i}\overset{iid}{\sim}\mathsf{N}(0,1)$,
$e_{i}\overset{iid}{\sim}\mathsf{N}(0,1)$, and $e_{i}$ and $\varepsilon_{i}$
are independent. An equivalent expression for these associations is
$\mu_{i}=\mu+\tau\varepsilon_{i},X_{i}=\mu+\tau\varepsilon_{i}+e_{i}$,
in which the data are directly linked to the unknown $\mu$. Since
the focus is on $\mu_{1}$, equations related to $\mu_{2},\ldots,\mu_{n}$
can be ignored. In the following two subsections we discuss the cases
with both known and unknown $\tau^{2}$.

\subsubsection{The case with a known $\tau^{2}$}

\label{subsec:normal_example_1}

This case corresponds to the motivating example presented in Section
\ref{sec:motivating_example}, and we are going to derive formula
\eqref{eq:motivating_example_interval} with $\sigma^{2}=1$. Since
$\tau$ is known, let $W_{i}=\tau\varepsilon_{i}+e_{i},i=1,2,\ldots,n$,
and then the system of associations $X_{i}=\mu+\tau\varepsilon_{i}+e_{i}$
can be rewritten as $\overline{X}=\mu+\overline{W}$ and $X_{i}-X_{1}=W_{i}-W_{1}$
for $i=2,\ldots,n$, where $\overline{X}=\frac{1}{n}\sum_{i=1}^{n}X_{i}$
and $\overline{W}=\frac{1}{n}\sum_{i=1}^{n}W_{i}$. Therefore, by
denoting $T(X)=\overline{X}$ and $H(X)=X_{(-1)}-X_{1}\mathbf{1}_{n-1}$,
where $X_{(-1)}=(X_{2},\ldots,X_{n})'$ and $\mathbf{1}_{n-1}$ is
a vector of all ones, the decomposition in equation \eqref{eq:CIM_decomposition}
is achieved. The associated auxiliary variable for $H(X)$ is $W_{H}=W_{(-1)}-W_{1}\mathbf{1}_{n-1}$,
where $W_{(-1)}=(W_{2},\ldots,W_{n})'$.

Next, we keep the following two associations $\overline{X}=\mu+\overline{W}$
and $\mu_{1}=\mu+\tau\varepsilon_{1}$, where $\overline{W}\sim\mathsf{P}_{\overline{W}|h}$
and $\varepsilon_{1}\sim\mathsf{P}_{\varepsilon_{1}|h}$ conditional
on $W_{H}=h\equiv H(x)$. The last step is to take $b(\overline{X},\mu_{1})=\overline{X}-\mu_{1}$,
and the final association equation is $b(\overline{X},\mu_{1})=W_{b}\coloneqq\overline{W}-\tau\varepsilon_{1}$.
It can be verified that the conditional distribution of $W_{b}$ given
$W_{H}=h$ is
\begin{equation}
W_{b}|\{W_{H}=h\}\sim\mathsf{N}\left(\frac{\tau^{2}}{1+\tau^{2}}(\bar{x}-x_{1}),\frac{n\tau^{2}+1}{n(\tau^{2}+1)}\right),\label{eq:normal_1_cdf}
\end{equation}
and the predictive random set \eqref{eq:random_set} can be constructed
accordingly. As a result, the conditional plausibility function for
$\mu_{1}$ is obtained as
\begin{equation}
\mathsf{cpl}_{T(x)|h}(\mu_{1})=2\Phi\left(-\left|\frac{\tau^{2}}{\tau^{2}+1}x_{1}+\frac{1}{\tau^{2}+1}\bar{x}-\mu_{1}\right|\left/\sqrt{\frac{n\tau^{2}+1}{n(\tau^{2}+1)}}\right.\right),\label{eq:normal_1_cpl}
\end{equation}
where $\Phi$ is the standard normal c.d.f., and hence the interval
estimator for $\mu_{1}$ is
\begin{equation}
C_{\alpha}(X)=\left(\frac{\tau^{2}}{\tau^{2}+1}X_{1}+\frac{1}{\tau^{2}+1}\overline{X}\right)\pm z_{\alpha/2}\sqrt{\frac{n\tau^{2}+1}{n(1+\tau^{2})}}.\label{eq:normal_1_interval}
\end{equation}

\subsubsection{The case with an unknown $\tau^{2}$}

\label{subsec:normal_example_2}

Similar to the previous case, the starting point is to decompose the
data associations into $T(X)$ and $H(X)$, which can be done in two
stages as described below. In the first stage, we keep the association
for $X_{1}$ and decompose $X_{(-1)}$ instead. Consider the ancillary
statistics $H_{i}(X)=(X_{i}-\overline{X}_{(-1)})/S_{(-1)}$ for $i=2,\ldots,n$,
where $\overline{X}_{(-1)}$ and $S_{(-1)}^{2}$ are the sample mean
and sample variance of $X_{(-1)}$. It is clear that $X_{(-1)}$ has
a one-to-one mapping to $(\overline{X}_{(-1)},S_{(-1)}^{2},H_{2}(X),\ldots,H_{n-1}(X))$.
Since marginally $X_{i}\overset{iid}{\sim}\mathsf{N}(\mu,\tau^{2}+1)$,
it is well known that $(\overline{X}_{(-1)},S_{(-1)}^{2})$ is a complete
sufficient statistic for $(\mu,\tau)$, and thus is independent of
$H_{i}(X)$ according to Basu's theorem. Therefore, conditioning on
$H_{i}(X)$ does not change the distribution of $(\overline{X}_{(-1)},S_{(-1)}^{2})$,
and we obtain the following four associations: \emph{(a)} $\mu_{1}=\mu+\tau\varepsilon_{1}$,
\emph{(b)} $X_{1}=\mu+\tau\varepsilon_{1}+e_{1}$, \emph{(c)} $\overline{X}_{(-1)}=\mu+\tilde{\tau}Z$
, and \emph{(d)} $S_{(-1)}^{2}=(\tau^{2}+1)M_{n-2}^{2}$, where $\tilde{\tau}=\sqrt{(\tau^{2}+1)/(n-1)}$,
$Z\sim\mathsf{N}(0,1)$, $M_{n-2}^{2}\sim\chi_{n-2}^{2}/(n-2)$, and
the auxiliary variables $\varepsilon_{1},e_{1},Z$, and $M_{n-2}^{2}$
are mutually independent. Equations \emph{(c)} and \emph{(d)} are
derived from the well-known facts that $\overline{X}_{(-1)}\sim\mathsf{N}(\mu,\tilde{\tau}^{2})$
and $(n-2)S_{(-1)}^{2}/(\tau^{2}+1)\sim\chi_{n-2}^{2}$.

Then in the second stage, we condition on the following equation,
as the auxiliary variable $W_{H}$ is known to follow a student $t$-distribution
with $n-2$ degrees of freedom:
\begin{equation}
H(X)\coloneqq\sqrt{\frac{n-1}{n}}\cdot\frac{X_{1}-\overline{X}_{(-1)}}{S_{(-1)}}=W_{H}\coloneqq\frac{\tau\varepsilon_{1}+e_{1}-\tilde{\tau}Z}{\sqrt{n}\tilde{\tau}M_{n-2}}\sim t_{n-2}.\label{eq:normal2_cond}
\end{equation}
As a result, we keep the associations $\mu_{1}=\mu+\tau\varepsilon_{1}$,
$\overline{X}_{(-1)}=\mu+\tilde{\tau}Z$, and $S_{(-1)}^{2}=(\tau^{2}+1)M_{n-2}^{2}$,
with $\varepsilon_{1}\sim\mathsf{P}_{\varepsilon_{1}|h},Z\sim\mathsf{P}_{Z|h}$,
and $M_{n-2}^{2}\sim\mathsf{P}_{M_{n-2}^{2}|h}$ conditional on $W_{H}=h\equiv H(x)$.
Obviously in this case $T(X)=(\overline{X}_{(-1)},S_{(-1)}^{2})$,
which combined with $H(X)$ completes the decomposition.

Next, by observing that $\overline{X}_{(-1)}-\mu_{1}$ is free of
$\mu$, we can take $b(T(X),\mu_{1})$ to be a function of $\overline{X}_{(-1)}-\mu_{1}$
and $S_{(-1)}^{2}$, so that the corresponding auxiliary variable
$W_{b}(\tau)$ is indexed by only one unknown parameter $\tau$. Specifically,
let
\begin{equation}
\begin{aligned}\tilde{\mu} & =\sqrt{\frac{n-1}{n}}h\left(\frac{n-2}{h^{2}+n-2}S_{(-1)}^{-1}-S_{(-1)}\right),\\
\tilde{\sigma}^{2} & =\max\left\{ n^{-\gamma},1-\frac{(n-1)(n-2)(n-3-h^{2})}{n(n-2+h^{2})^{2}}S_{(-1)}^{-2}\right\} ,\ \gamma\in(0,\tfrac{1}{2}),
\end{aligned}
\label{eq:normal_2_mu_sigma}
\end{equation}
and then define $b(T(X),\mu_{1})=(\overline{X}_{(-1)}-\mu_{1}-\tilde{\mu})/\tilde{\sigma}$,
where $\tilde{\mu}$ and $\tilde{\sigma}$ are chosen such that $\mathbb{E}(W_{b}(\tau)|W_{H}=h)=0$
and that $W_{b}(\tau)|\{W_{H}=h\}\overset{d}{\rightarrow}\mathsf{N}(0,1)$.
These two conditions ensures that $W_{b}(\tau)$ will be gradually
free of $\tau$ when $n$ is large. Next, let $F_{W_{b}(\tau)|h}$
be the c.d.f. of $W_{b}(\tau)$ given $W_{H}=h$, and we can show
that
\begin{equation}
F_{W_{b}(\tau)|h}(s)=\int_{0}^{+\infty}\Phi\left(\frac{s\sqrt{\max\left\{ n^{-\gamma},1-c_{1}\omega/x\right\} }-c_{2}\sqrt{\omega}\left(\sqrt{x}-c_{3}/\sqrt{x}\right)}{\sqrt{1-\omega(n-1)/n}}\right)g(x)\mathrm{d}x,\label{eq:normal_2_cdf}
\end{equation}
where $\omega=(1+\tau^{2})^{-1}$, $c_{1}=(n-2)(n-3-h^{2})/\{n(h^{2}+n-2)\}$,
$c_{2}=(n-1)h/\sqrt{n(h^{2}+n-2)}$, $c_{3}=(n-2)/(n-1)$, and $g$
is the p.d.f. of $\chi_{n-1}^{2}/(n-1)$.

Finally, let $\underline{F}(s)=\inf_{\omega\in(0,1)}F_{W_{b}(\tau)|h}(s)$
and $\overline{F}(s)=\sup_{\omega\in(0,1)}F_{W_{b}(\tau)|h}(s)$,
both computable using numerical methods, and we can show that
\[
\mathsf{cpl}_{T(x)|h}(\mu_{1})=\min\left\{ 1,2\left[1-\underline{F}\left(\frac{x_{1}-\mu_{1}-\tilde{\mu}}{\tilde{\sigma}}\right)\right],2\overline{F}\left(\frac{x_{1}-\mu_{1}-\tilde{\mu}}{\tilde{\sigma}}\right)\right\} ,
\]
and that
\[
C_{\alpha}(X)=\left(\overline{X}_{(-1)}-\tilde{\mu}-\underline{F}^{-1}(1-\alpha/2)\tilde{\sigma},\ \overline{X}_{(-1)}-\tilde{\mu}-\overline{F}^{-1}(\alpha/2)\tilde{\sigma}\right).
\]

\subsection{The Poisson Hierarchical Model}

\label{subsec:poisson_example}

The Poisson hierarchical model is useful for analyzing discrete data
such as counts. Assume that given parameters $\lambda_{i}>0$, the
observed data $X=(X_{1},\ldots,X_{n})'$ satisfy $X_{i}|\lambda_{i}\sim\mathsf{Pois}(\lambda_{i}t_{i}),i=1,\ldots,n$,
where $t_{i}>0$ are known constants. In real-world problems, $\lambda_{i}$
can be interpreted, for example, as the rate of events in unit time,
and $t_{i}$ is the length of the time window. It is also assumed
that $\lambda_{i}$'s follow a common prior, $\lambda_{i}\overset{iid}{\sim}\gamma\mathsf{Gamma}(s)$,
where $s$ is a known shape parameter and $\gamma$ is an unknown
scale parameter. In this setting the parameter of interest is $\lambda_{1}$.
This model can also be expressed using the formulation in Section
\ref{subsec:model_specification}:

\begin{center}
\begin{tabular}{ll}
\toprule 
\addlinespace[0.1em]
Sampling model & $X|(\tilde{\theta},\theta^{*})\sim\prod_{i}\mathsf{Pois}(\lambda_{i}t_{i})$\tabularnewline\addlinespace[0.1em]
\addlinespace[0.1em]
Partial prior & $\tilde{\theta}=(\lambda_{1},\lambda_{2},\ldots,\lambda_{n})$, $\tilde{\theta}|\theta^{*}\sim\prod_{i}\gamma\mathsf{Gamma}(s)$\tabularnewline\addlinespace[0.1em]
\addlinespace[0.1em]
Component without prior & $\theta^{*}=\gamma$\tabularnewline\addlinespace[0.1em]
\addlinespace[0.1em]
Parameter of interest & $\eta=\lambda_{1}$\tabularnewline\addlinespace[0.1em]
\bottomrule
\end{tabular}
\par\end{center}

For this Poisson hierarchical model, the data associations and prior
associations are given by $X_{i}=F_{\lambda_{i}t_{i}}^{-1}(U_{i})$
and $\lambda_{i}=\gamma V_{i}$, respectively, with $i=1,\ldots,n$.
$F_{\lambda}^{-1}$ is the generalized inverse c.d.f. of the Poisson
distribution with mean $\lambda$, $U=(U_{1},\ldots,U_{n})'\overset{iid}{\sim}\mathsf{Unif}(0,1),V=(V_{1},\ldots,V_{n})'\overset{iid}{\sim}\mathsf{Gamma}(s)$,
and $U$ and $V$ are independent. After plugging prior associations
into data associations and ignoring irrelevant parameters, the following
association equations are kept without loss of information:
\begin{equation}
\lambda_{1}=\gamma V_{1},\ \text{and}\ X_{i}=F_{\gamma V_{i}t_{i}}^{-1}(U_{i}),\ i=1,\ldots,n.\label{eq:poisson_association}
\end{equation}

A fundamental difference between this Poisson model and the normal
model studied earlier is that, due to the discreteness of $X_{i}$
and the heterogeneity of the $t_{i}$ values, it is improbable to
find a non-trivial function $H(x)$ such that the distribution of
$H(X)$ is free of $\gamma$. This is an example that the decomposition
\eqref{eq:CIM_decomposition} is not available, and hence we trivially
take $H(X)=1$ and $T(X)=X$. As a result, the next step is to seek
the $b$ function in \eqref{eq:final_association} such that $b(T(X),\lambda_{1})$
only weakly relies on $\lambda_{1}$. The idea is as follows.

First fix $\lambda_{1}$ to its true realization, and assume that
an approximation of $\lambda_{1}$, denoted by $\hat{\lambda}_{1}$,
is given. We then require that $b(x,\hat{\lambda}_{1})=0$ and $\partial b/\partial\lambda_{1}|_{\lambda_{1}=\hat{\lambda}_{1}}=0$,
which indicates that $b$ is almost free of $\lambda_{1}$ in a neighborhood
of $\hat{\lambda}_{1}$. If $\hat{\lambda}_{1}$ is chosen to be the
MLE, then $b$ is obtained as the likelihood ratio function, \emph{i.e.},
$b(x,\lambda_{1})=\ell(\hat{\lambda}_{1};x)-\ell(\lambda_{1};x)$,
where $\ell(\lambda_{1};x)=\log f(x|\lambda_{1})$ is the log density
function of $X$ conditional on $\lambda_{1}$, and $\hat{\lambda}_{1}=\hat{\lambda}_{1}(x)=\arg\max_{\lambda_{1}}\ell(\lambda_{1};x)$.

Note that in the associations \eqref{eq:poisson_association}, $X_{i}$
can also be written as $X_{i}=F_{\lambda_{1}t_{i}V_{i}/V_{1}}^{-1}(U_{i})$
with respect to $\lambda_{1}$, and we express it as $X=a(\lambda_{1},U,V)$
for simplicity. Therefore, given the $b$ function, the final association
\eqref{eq:final_association} then becomes $b(X,\lambda_{1})=W_{b}(\lambda_{1})$,
where the auxiliary variable $W_{b}(\lambda_{1})$ is defined by
\[
W_{b}(\lambda_{1})=\ell(\hat{\lambda}_{1}(a(\lambda_{1},U,V));a(\lambda_{1},U,V))-\ell(\lambda_{1};a(\lambda_{1},U,V)).
\]
Let $G_{\lambda_{1}}$ be the c.d.f. of $W_{b}(\lambda_{1})$ conditional
on $\lambda_{1}$, and then the unconditional plausibility function
for $\lambda_{1}$ is $\mathsf{pl}_{x}(\lambda_{1})=1-G_{\lambda_{1}}(b(x,\lambda_{1}))$.
Finally, the interval estimator for $\lambda_{1}$ is obtained by
inverting the plausibility function, \emph{i.e.,} $C_{\alpha}(x)=\{\tilde{\lambda}:\mathsf{pl}_{x}(\tilde{\lambda})\ge\alpha\}$.
The computation details are given in Appendix \ref{subsec:poisson_computation}.

The choice of the $b$ function is not unique, and the one used here
is inspired by \citet{martin2015plausibility}. Due to the choice
of $T(X)=X$, Theorem \ref{thm:efficiency} no longer applies to this
case, but the simulation result in Section \ref{sec:simulation} suggests
that the interval estimator derived in this section is indeed very
efficient. Also, it is worth mentioning that the validity property
always holds regardless of the choice of $b$.

\subsection{The Binomial Rates-Difference Model}

\label{subsec:binomial_example}

The last binomial rates-difference model is motivated by a clinical
trial study \citep{xie2013incorporating}. It can be described as
follows. Assume that two independent binomial samples, $X$ and $Y$,
were collected with $X\sim\mathsf{Bin}(m,p_{1})$, and $Y\sim\mathsf{Bin}(n,p_{2})$.
The available prior information is on the difference of the success
rates, $\delta\coloneqq p_{1}-p_{2}\sim\pi$, and the task is to make
inference about $\delta$. For this model, we have

\begin{center}
\begin{tabular}{ll}
\toprule 
\addlinespace[0.1em]
Sampling model & $X|(\tilde{\theta},\theta^{*})\sim\mathsf{Bin}(m,p_{1})$, $Y|(\tilde{\theta},\theta^{*})\sim\mathsf{Bin}(n,p_{2})$\tabularnewline\addlinespace[0.1em]
\addlinespace[0.1em]
Partial prior & $\tilde{\theta}=\delta\coloneqq p_{1}-p_{2}$, $\tilde{\theta}|\theta^{*}\sim\pi$\tabularnewline\addlinespace[0.1em]
\addlinespace[0.1em]
Component without prior & $\theta^{*}=p_{1}+p_{2}$\tabularnewline\addlinespace[0.1em]
\addlinespace[0.1em]
Parameter of interest & $\eta=\delta$\tabularnewline\addlinespace[0.1em]
\bottomrule
\end{tabular}
\par\end{center}

Obviously, the data association equations of this model are $X=F_{m,p_{1}}^{-1}(U_{1})$
and $Y=F_{n,p_{2}}^{-1}(U_{2})$, and the prior association is $\delta=U$,
where $F_{k,p}^{-1}$ is the generalized inverse c.d.f. of $\mathsf{Bin}(k,p)$.
The auxiliary variables $U_{1},U_{2}\overset{iid}{\sim}\mathsf{Unif}(0,1)$,
$U\sim\pi$, and $U_{1},U_{2}$, and $U$ are independent. To simplify
the notations, $p_{1}$ and $p_{2}$ are re-parameterized as $\delta=p_{1}-p_{2}$
and $\tau=p_{1}+p_{2}$. Since $p_{1}$ and $p_{2}$ must lie in $[0,1]$,
$\tau$ is further written as $\tau=1+(1-|\delta|)\omega$ to guarantee
the range, where $\omega\in(-1,1)$ is an unknown quantity. As a result,
$p_{1}=p_{1}(\delta,\omega)=\{1+\delta+(1-|\delta|)\omega\}/2$ and
$p_{2}=p_{2}(\delta,\omega)=\{1-\delta+(1-|\delta|)\omega\}/2$ are
functions of the new parameters $\delta$ and $\omega$.

Similar to the association steps of previously studied models, we
first plug the prior association into the data association, resulting
in
\[
X=F_{m,p_{1}(U,\omega)}^{-1}(U_{1}),\ Y=F_{n,p_{2}(U,\omega)}^{-1}(U_{2}),\ \text{and}\ \delta=U.
\]
Again due to the discreteness of $X$ and $Y$, it is unlikely to
find a function $H(X,Y)$ such that its distribution is free of $\omega$
, so the goal is to seek the $b$ function as in the Poisson model.
Like in the Poisson case, we first find an approximation $\hat{\delta}$
to $\delta$, and then solve the functional equations $b(x,y,\hat{\delta})=0$
and $\partial b/\partial\delta|_{\delta=\hat{\delta}}=0$.

However, this model has two significant differences from the Poisson
case: first, $\delta$ has a genuine prior $\delta\sim\pi$, and second,
there is one more unknown parameter $\omega$. Our proposal here is
to use the maximum a posteriori estimator for $\delta$ as the approximation,
derived as follows: let $f(x,y,\delta;\omega)$ be the joint density
function of $(X,Y,\delta)$ and define $\ell(\delta,\omega;x,y)=\log f(x,y,\delta;\omega)$.
$\hat{\delta}$ is then obtained as $(\hat{\delta},\hat{\omega})=\arg\max_{\delta,\omega}\ell(\delta,\omega;x,y)$.
With $\hat{\delta}=\hat{\delta}(x,y)$ and $\hat{\omega}=\hat{\omega}(x,y)$,
$b$ can be solved as
\begin{equation}
b(x,y,\delta)=\ell(\hat{\delta}(x,y),\hat{\omega}(x,y);x,y)-\ell(\delta,\hat{\omega}_{\delta}(x,y);x,y),\label{eq:binomial_b}
\end{equation}
where $\hat{\omega}_{\delta}(x,y)=\arg\max_{\omega}\ell(\delta,\omega;x,y)$.

In the last step, the final association is $b(X,Y,\delta)=W_{b}(\omega)$,
where $W_{b}(\omega)$ is obtained by replacing $(x,y,\delta)$ with
$\left(F_{m,p_{1}(U,\omega)}^{-1}(U_{1}),F_{n,p_{2}(U,\omega)}^{-1}(U_{2}),U\right)$
in \eqref{eq:binomial_b}. Let $G_{\omega}$ denote the c.d.f. of
$W_{b}(\omega)$, and define $\underline{G}(s)=\inf_{\omega\in(-1,1)}G_{\omega}(s)$,
and then the unconditional plausibility function for $\delta$ is
$\mathsf{pl}_{x,y}(\delta)=1-\underline{G}(b(x,y,\delta))$, with
the interval estimator $C_{\alpha}(X,Y)$ defined by $C_{\alpha}(x,y)=\{\tilde{\delta}:\mathsf{pl}_{x,y}(\tilde{\delta})\ge\alpha\}.$
The computation details are given in Appendix \ref{subsec:binomial_computation}.

\section{Simulation Study}

\label{sec:simulation}

In this section we conduct several simulation studies to compare Partial
Bayes solutions with other existing methods such as Empirical Bayes
and Confidence Distribution approaches. Specifically, given the observed
data from a model and the parameter of interest, each method computes
an interval estimator for the parameter. Data are simulated 10,000
times in order to calculate the empirical coverage percentage and
the mean interval width for all the methods compared. The nominal
coverage rate is set to 95\% for all experiments. In the following
part, the three popular models studied in Section \ref{sec:popular_models}
are considered.

\paragraph{The Normal Hierarchical Model}

\begin{figure}[h]
\begin{centering}
\includegraphics[width=0.95\textwidth]{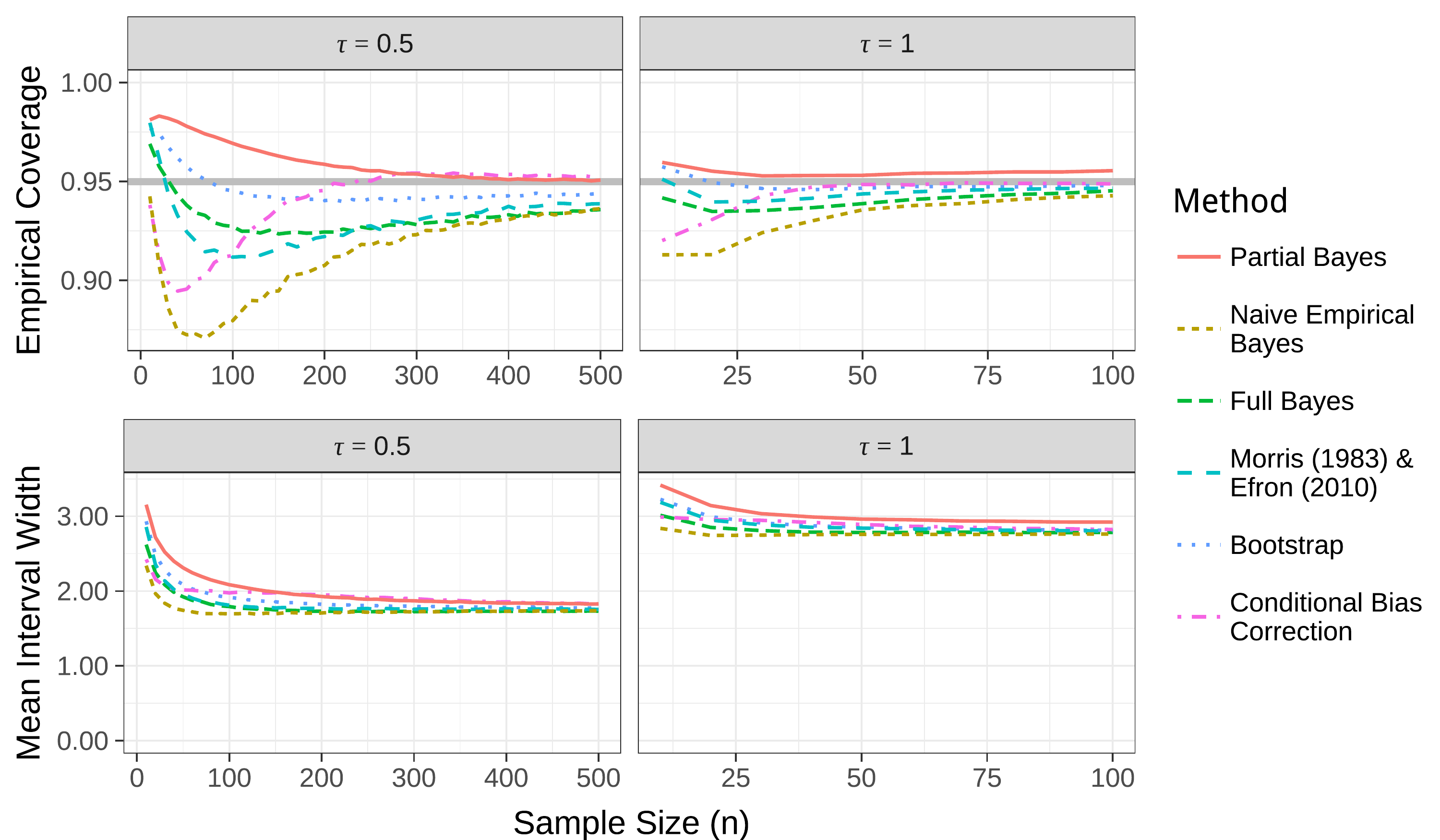}
\par\end{centering}
\caption{The empirical coverage percentage (the top two panels) and mean interval
width (the bottom two panels) for $\mu_{1}$ in the normal hierarchical
model with an increasing sample size $n$ and two parameter settings,
among 10,000 simulation runs. For all the methods compared, only the
Partial Bayes solution guarantees the nominal coverage rate for all
$n$. \label{fig:simulation_normal}}

\end{figure}

The normal hierarchical model in Section \ref{subsec:normal_example}
is extremely popular in literature. In this experiment the Partial
Bayes solution is compared with the naive Empirical Bayes and other
improved methods, including the full Bayes method with flat prior
\citep{deely1981bayes}, the approach used by \citet{morris1983parametric}
and \citet{efron2010large}, the Bootstrap method \citep{laird1987empirical},
and the Conditional Bias Correction method \citep{carlin1990approaches}.
In this model, both hyper-parameters $\mu$ and $\tau^{2}$ are assumed
to be unknown, with the same setting in \citet{laird1987empirical}:
the true $\mu$ is fixed to 0, and two values of $\tau$, 0.5 and
1, are considered. For the Partial Bayes solution, the $\gamma$ constant
in \eqref{eq:normal_2_mu_sigma} is fixed to be $\frac{1}{3}$. The
results of the empirical coverage percentage and the mean interval
width for different methods are summarized in Figure \ref{fig:simulation_normal}.

It is obvious in Figure \ref{fig:simulation_normal} that among all
the methods compared, only the Partial Bayes solution achieves the
nominal coverage rate for all sample sizes. In terms of interval width,
the Partial Bayes solution has wider interval estimates than other
methods, due to the guarantee of coverage rate; however, as the sample
size increases, the gaps between different methods become smaller
and smaller, indicating that all methods are efficient asymptotically.

\paragraph{The Poisson Hierarchical Model}

\begin{figure}[h]
\begin{centering}
\includegraphics[width=0.95\textwidth]{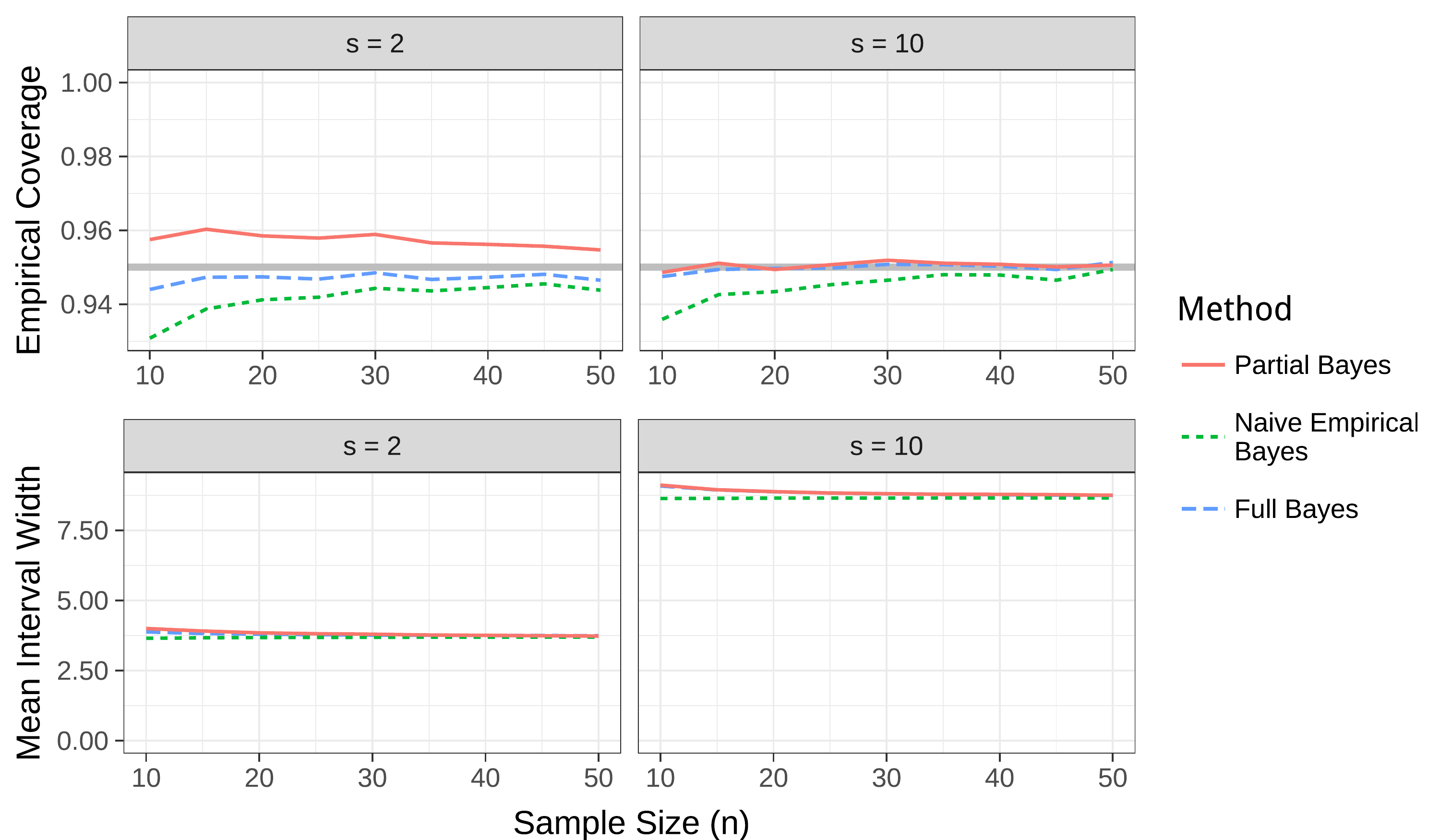}
\par\end{centering}
\caption{The empirical coverage percentage (the top two panels) and mean interval
width (the bottom two panels) for $\lambda_{1}$ in the Poisson hierarchical
model with an increasing sample size $n$ and two parameter settings,
among 10,000 simulation runs. Three different solutions are compared,
showing that the Partial Bayes solution guarantees the nominal coverage
rate. \label{fig:simulation_poisson}}
\end{figure}

The second simulation experiment is for the Poisson hierarchical model
discussed in Section \ref{subsec:poisson_example}. For simplicity,
we set all the $t_{i}'s$ to be 1, and fix the true value of $\theta$
to be 1. Two different values of $s$, $s=2,10$, and a sequence of
sample sizes, $n=10,15,\ldots,50$, are considered. There are fewer
existing results for the Poisson model than the normal one, and here
the Partial Bayes solution is compared with the naive Empirical Bayes
and full Bayes approaches, with the results illustrated in Figure
\ref{fig:simulation_poisson}.

The pattern of the simulation results is very similar to that of the
normal model. As expected, the other two solutions have narrower interval
estimates than the Partial Bayes solution, but they do not preserve
the nominal coverage rate. In contrast, the Partial Bayes solution
has coverage percentages above 95\%, and its interval width is getting
close to the other two when sample size increases. The simulation
result again verifies both the exactness and the efficiency of the
Partial Bayes solution.

\paragraph{The Binomial Rates-Difference Model}

\begin{figure}[h]
\begin{centering}
\includegraphics[width=0.95\textwidth]{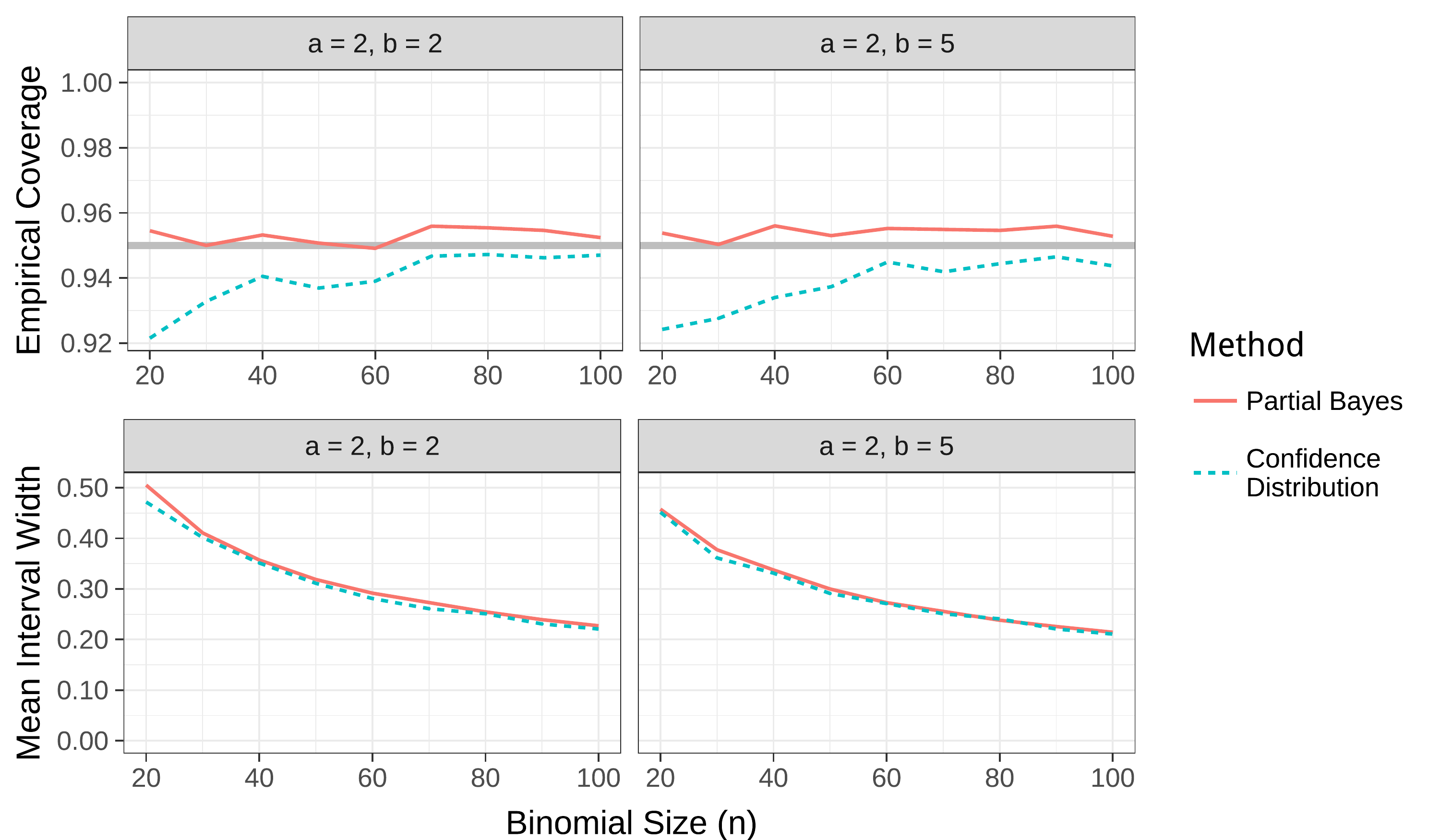}
\par\end{centering}
\caption{The empirical coverage percentage (the top two panels) and mean interval
width (the bottom two panels) for $\delta$ in the binomial rates-difference
model with an increasing binomial size $n$ and two parameter settings,
among 10,000 simulation runs. Partial Bayes and Confidence Distribution
solutions are compared, showing that the Partial Bayes solution guarantees
the nominal coverage for all $n$. \label{fig:simulation_binomial}}
\end{figure}

In the last experiment we consider the binomial model studied in Section
\ref{subsec:binomial_example}. The prior of $\delta\equiv p_{1}-p_{2}$
is chosen to have the same distribution as $2\beta-1$ with $\beta\sim\mathsf{Beta}(a,b)$
for some known value of $(a,b)$. This choice of prior guarantees
that the support of $\pi(\delta)$ is $[-1,1]$. For each simulated
$\delta$, the value of $\tau\equiv p_{1}+p_{2}$ is created as $\tau=1+(1-|\delta|)\omega$
with $\omega\sim\mathsf{Unif}(-1,1)$. Then the corresponding true
values of $p_{1}$ and $p_{2}$ used to simulate the data can be determined
accordingly. Two settings of prior distribution parameters, $(a,b)=(2,2)$
and $(2,5)$, and a sequence of binomial sizes, $m=n=20,30,\ldots,100$,
are considered. Since the typical Empirical Bayes methods do not apply
to this problem, in Figure \ref{fig:simulation_binomial} we give
the results of Partial Bayes and Confidence Distribution solutions.

Similar to the Empirical Bayes solutions in the previous two simulation
studies, Confidence Distribution does not possess the desired coverage,
while Partial Bayes provides exact inference results. This is because
the Confidence Distribution method for this model relies on large
sample theory, and may not work well for small samples. The interval
width of the Partial Bayes solution is slightly wider than that of
the Confidence Distribution method, but the difference is only tiny;
as expected, the width will decrease as sample size increases, which
again indicates the efficiency.

\section{Application}

\label{sec:application}

In this section we apply the Partial Bayes model to a dataset of National
Basketball Association (NBA) games. In basketball competitions, a
three-point shot, if made, rewards the highest score in one single
attempt. Therefore, as the game comes to an end, three-point shots
are more valuable for a team that has very limited offensive possessions
and needs to overcome the deficit in score. When the game is decided
by the last possession, a three-point shot is usually beneficial or
even necessary for such teams, and the choice of player that will
make the attempt is crucial to the outcome of the game.

Typically, the player to be chosen should have the highest success
rate of three-point shots, and historical data can be used to evaluate
each player's performance. If $X_{i}$ is the number of three-point
shots made in $n_{i}$ attempts by player $i$, then usually $X_{i}$
can be modeled by a binomial distribution $\mathsf{Bin}(n_{i},p_{i})$
or a Poisson distribution $\mathsf{Pois}(n_{i}p_{i})$, where $p_{i}$
stands for the success rate. In this application we choose the latter
one for simplicity. Given this model, a classical point estimator
for $p_{i}$ is $\hat{p}_{i}=X_{i}/n_{i}$, and a $100(1-\alpha)\%$
frequentist confidence interval for $p_{i}$ is $\left(G_{X_{i}}\left(\frac{\alpha}{2}\right)/n_{i},G_{X_{i}+1}\left(1-\frac{\alpha}{2}\right)/n_{i}\right)$,
where $G_{s}(\cdot)$ is the c.d.f. of the $\mathsf{Gamma}(s)$ distribution.

If additional information is available, for example $p_{i}$'s are
assumed to follow a common prior distribution $\pi(p)$, then the
efficiency of the inference can be improved by incorporating this
prior. This assumption is sensible since the players are in the same
team or league, and they are expected to share some common characteristics.
By combining the two sources of information --- player's own historical
statistics, and those of other players in the team or league ---
a more fair evaluation of players' performance could then be obtained.
In what follows, we analyze the three-point shot data obtained from
the official NBA website. We first select three players from each
team that have the highest three-point goal success rates during the
2015-2016 regular season, and then retrieve the data from each player's
last ten games within that season. The number of three-point shots
made ($X_{i}$) and attempted ($n_{i}$) for each player are computed
from this dataset.

To take the prior information into account, we first use the Empirical
Bayes method to analyze this dataset similar to the analysis in \citet{efron1975data}
for baseball games, but with a Poisson model instead of a normal one.
The $p_{i}$'s are assumed to follow a common exponential prior $exp(\theta)$,
where $\theta>0$ stands for the mean. The MLE of $\theta$ is obtained
as $\hat{\theta}=0.410$ using the marginal distribution of $X_{i}$.
As a result, the point estimator for $p_{i}$ is taken to be the posterior
mean $(X_{i}+1)/(\hat{\theta}^{-1}+n_{i})$, and the approximate $100(1-\alpha)\%$
Bayesian credible interval is $\left(G_{X_{i}+1}\left(\frac{\alpha}{2}\right)/(\hat{\theta}^{-1}+n_{i}),G_{X_{i}+1}\left(1-\frac{\alpha}{2}\right)/(\hat{\theta}^{-1}+n_{i})\right)$.

\begin{figure}[h]
\begin{centering}
\includegraphics[width=0.8\textwidth]{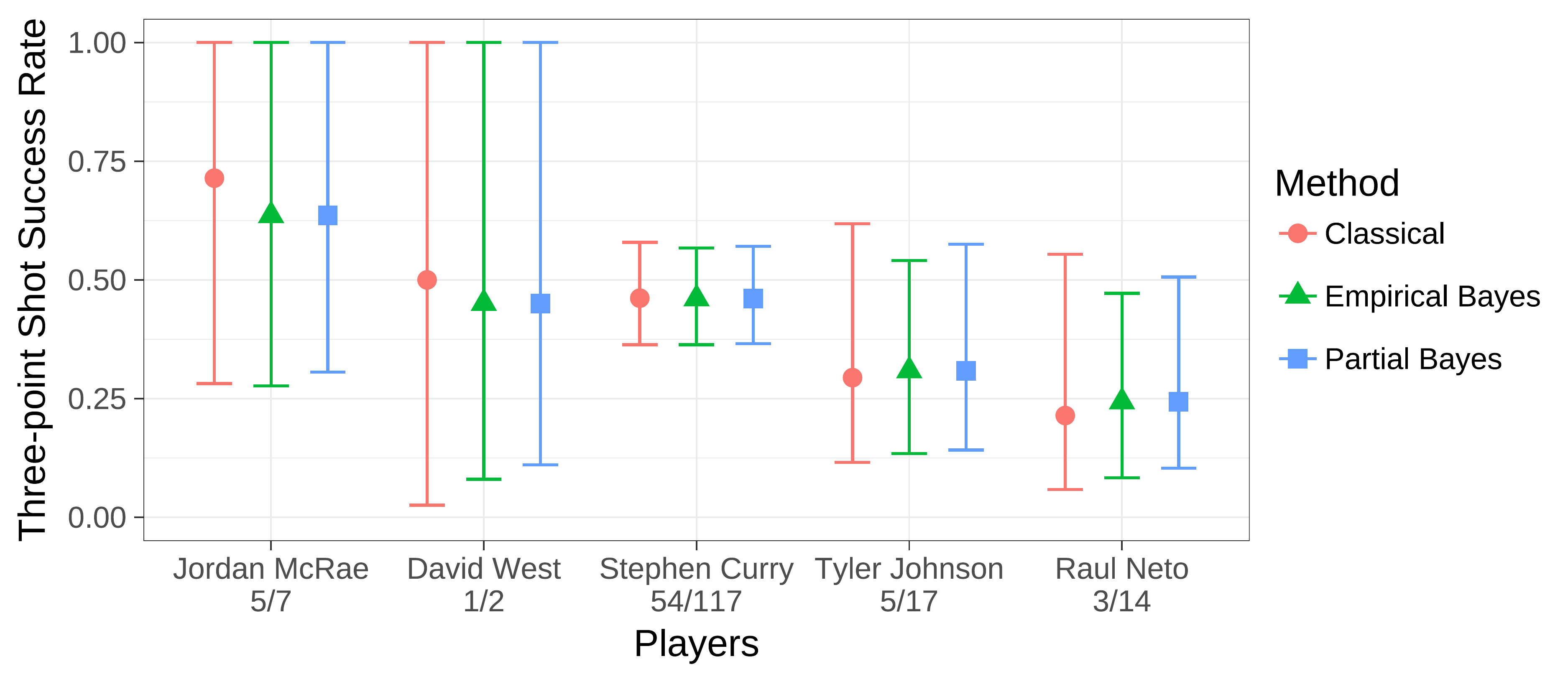}
\par\end{centering}
\caption{Comparing three methods for analyzing three-point shot success rates
on five representative players among the ninety players studied. Numbers
of three-point shots made and attempted are displayed under players'
names. The error bars and the dots stand for the 90\% interval estimates
and the point estimates respectively. The three different shapes of
dots represent the three inference methods. \label{fig:application}}
\end{figure}

Finally, the Partial Bayes model in Section \ref{subsec:poisson_example}
is used to derive an interval estimator for $p_{i}$, and the point
estimator is chosen as the value of $p_{i}$ that maximizes $\mathsf{pl}_{x}(p_{i})$.
The comparison of the three methods mentioned above is shown in Figure
\ref{fig:application} for five representative players.

Among these five players, Jordan McRae and David West are examples
of players with high success rates but few number of shot attempts.
It is clear that both Empirical Bayes and Partial Bayes results shrink
the classical point estimates towards the grand mean, as an effect
of combining individual and league information. To the opposite, for
players below the average, such as Tyler Johnson and Raul Neto, their
success rates are lifted by a small percentage. Stephen Curry, as
a third case, is almost unaffected by the shrinkage. This is because
he made a large number of shot attempts, so that his personal performance
dominates the overall estimate. It is worth noting that David West
has a higher point estimate of success rate than Stephen Curry in
the classical method, but their rankings are reversed in Empirical
Bayes and Partial Bayes methods.

The comparison of the three methods also highlights the advantage
of the Partial Bayes method. It is known that the classical confidence
interval is exact, but is wider than that of the other two methods.
The Empirical Bayes solution is more efficient, but theoretically
it is only approximate. The Partial Bayes solution, in contrast, combines
the advantages of the other two methods, providing both exact and
efficient inference results. This example hence suggests that the
Partial Bayes model framework is useful for real-life data analysis
tasks.

\section{Conclusion and Discussion}

\label{sec:conclusion}

This article considers the statistical inference for Partial Bayes
problems, \emph{i.e.}, Bayesian models without fully-specified prior
distributions. We have developed a general model framework for studying
such problems, and have provided theoretical justification for both
the exactness and the efficiency of the inference results. Compared
with other existing methodologies dealing with partial prior information,
such as Empirical Bayes and Confidence Distribution, our proposed
method has shown superior performance.

Indeed, statisticians and scientists do care about exact inference
for such useful models. For example, pioneering work in the Empirical
Bayes literature, such as \citet{morris1983parametric,laird1987empirical,carlin1990approaches},
has revealed the fact that Empirical Bayes estimators could underestimate
the uncertainty, and these authors all emphasized the importance of
providing exact inference for such problems. To some extent our discussion
sheds new light on this issue and shows promising results. From this
perspective, Partial Bayes models are powerful extensions to conventional
Bayesian models, as they allow for more flexibility on the prior specifications,
and meanwhile avoid sacrificing the exactness of inference. As a result,
they can be used to combine different types of information for which
other existing methods are difficult.

Of course, ``There is no such thing as a free lunch.'' The exact
and efficient inference for Partial Bayes problems is very useful
yet challenging. As has been illustrated by the three examples models,
the construction of the interval estimators can sometimes be quite
technical and non-trivial. Also, similar to the hierarchical Bayesian
models, the computational cost for Partial Bayes solutions may be
massive when the model structure is complex. Despite all these obstacles,
we believe that the Partial Bayes model framework is useful in real
data analysis, and we expect that more research along this direction
can be fruitful, as far as exact and efficient probabilistic inference
concerns.

\appendix

\section{Appendix}

\subsection{Proof of Theorem \ref{thm:cond_validity}}

Let $Q_{h}(w)=P_{\mathcal{S}_{h}}(w\notin\mathcal{S}_{h})$, and then
for any $(x,w,\tilde{\eta})$ such that $b(T(x),\tilde{\eta})=w$,
\begin{align}
\mathsf{cpl}_{T(x)|h}(\tilde{\eta}) & =1-P_{\mathcal{S}_{h}}(\Theta_{T(x)}(\mathcal{S}_{h})\subseteq(-\infty,\tilde{\eta})\cup(\tilde{\eta},+\infty))\nonumber \\
 & =1-P_{\mathcal{S}_{h}}(\tilde{\eta}\notin\Theta_{T(x)}(\mathcal{S}_{h}))\nonumber \\
 & =1-P_{\mathcal{S}_{h}}(w\notin\mathcal{S}_{h})\equiv1-Q_{h}(w).\label{eq:cpl_Q}
\end{align}
Therefore,
\begin{equation}
\mathsf{cpl}_{T(X)|h}(\eta)\ge\alpha\Leftrightarrow Q_{h}(W_{b}(\theta^{*}))\le1-\alpha.\label{eq:cpl_Qh}
\end{equation}
First fix $\theta^{*}$, and let $\mathsf{P}_{T(X),\eta|H(X)=h}$
denote the probability measure of $(T(X),\eta)$ given $H(X)=h$,
and then we see that $\mathsf{P}_{T(X),\eta|H(X)=h}\equiv\mathsf{P}_{W_{T},V_{\eta}|h}$.
As a result, we apply the probability measure $\mathsf{P}_{W_{T},V_{\eta}|h}$
on both sides of \eqref{eq:cpl_Qh}, obtaining
\[
P_{T(X),\eta|H(X)=h}\left(\mathsf{cpl}_{T(X)|h}(\eta)\ge\alpha\right)=P_{W_{b}(\theta^{*})|h}\left(Q_{h}(W_{b}(\theta^{*}))\le1-\alpha\right).
\]
The validity of $\mathcal{S}_{h}$ implies $P_{W_{b}(\theta^{*})|h}\left(Q_{h}(W_{b}(\theta^{*}))\ge1-\alpha\right)\le\alpha$
for any $\theta^{*}$. Therefore,
\begin{equation}
P_{X,\eta|H(X)}(C_{\alpha}(X)\ni\eta|H(X)=h)=P_{T(X),\eta|H(X)=h}\left(\mathsf{cpl}_{T(X)|h}(\eta)\ge\alpha\right)\ge1-\alpha.\label{eq:proof_validity}
\end{equation}
Note that \eqref{eq:proof_validity} is true for any fixed $\theta^{*}$,
so it also holds with $\theta^{*}\sim\pi^{*}(\theta^{*})$, for any
$\pi^{*}(\theta^{*})$.

\subsection{Proof of Theorem \ref{thm:optimality}}

Similar to \eqref{eq:cpl_Q}, we have $\text{\ensuremath{\mathsf{cpl}}}_{T(x)|h,t}(\tilde{\eta})\ge\alpha\Leftrightarrow Q_{h,t}(w)\le1-\alpha$,
where $(x,w,\tilde{\eta})$ satisfies $b(T(x),\tilde{\eta})=w$, and
$Q_{h,t}(w)=P_{\mathcal{S}_{h,t}}(w\notin\mathcal{S}_{h,t})$. Fixing
$t\equiv T(x)$, $\eta\mapsto b(t,\eta)$ is one-to-one by definition,
so the mapping must be monotone. Without loss of generality we assume
$b(t,\eta)$ is increasing in $\eta$, since otherwise we can use
$-b$ in place of $b$.

Let $Z_{\eta}=a_{\eta}(U,V_{\eta})$, and then it can be shown that
\begin{align*}
F_{W_{b}|h,t}(w) & =P_{W_{b}|h,t}\left(W_{b}\le w|W_{H}=h,Z_{T}=t\right)\\
 & =P_{Z_{T},Z_{\eta}|h,t}\left(b(Z_{T},Z_{\eta})\le b(t,\tilde{\eta})|W_{H}=h,Z_{T}=t\right)\\
 & =P_{Z_{\eta}|h,t}\left(b(t,Z_{\eta})\le b(t,\tilde{\eta})|W_{H}=h,Z_{T}=t\right)\\
 & =P_{U,V_{\eta}|h,t}\left(a_{\eta}(U,V_{\eta})\le\tilde{\eta}|W_{H}=h,Z_{T}=t\right).
\end{align*}
By the definition of the decomposition in \eqref{eq:CIM_decomposition},
$W_{H}=h,a_{T}(\theta^{*},W_{T})=t\Leftrightarrow a(\theta^{*},W)=x$,
and hence $W_{H}=h,Z_{T}=t\Leftrightarrow a(U,W)=x$. Also it is clear
from the association equations that $(\theta^{*},\eta,X)\equiv\left(U,a_{\eta}(U,V_{\eta}),a(U,W)\right)$,
so we have $F_{W_{b}|h,t}(w)=P_{\eta|X=x}(\eta\le\tilde{\eta}|X=x)=F_{\eta|x}(\tilde{\eta})$.

Finally, let $u=F_{W_{b}|h,t}(w)$. Since
\[
\mathcal{S}_{h,t}=\left\{ F_{W_{b|h,t}}^{-1}(u'),u'\in(0,1):\vert u'-0.5\vert<\vert U_{\mathcal{S}}-0.5\vert\right\} ,\ U_{\mathcal{S}}\sim\mathsf{Unif}(0,1),
\]
we have
\begin{align*}
Q_{h,t}(w) & =P_{\mathcal{S}_{h,t}}(w\notin\mathcal{S}_{h,t})=P_{U_{\mathcal{S}}}\left(|u-0.5|\ge|U_{\mathcal{S}}-0.5|\right)=|1-2u|=|1-2F_{\eta|x}(\tilde{\eta})|,
\end{align*}
and hence $\text{\ensuremath{\mathsf{cpl}}}_{T(x)|h,t}(\tilde{\eta})\ge\alpha\Leftrightarrow Q_{h,t}(w)\le1-\alpha\Leftrightarrow\alpha/2\le F_{\eta|x}(\tilde{\eta})\le1-\alpha/2$.

\subsection{Proof of Theorem \ref{thm:efficiency}}

We first show that $Z_{T_{n}}\overset{P}{\rightarrow}U$ and $W_{b_{n}}\overset{P}{\rightarrow}V_{\eta}$
under conditions \emph{(a)} and \emph{(b)}. Let $\mathsf{P}_{U}$
be the probability measure of $U$. Since $U$ and $W_{T_{n}}$ are
independent, we have that for any $\varepsilon>0$, $P(|Z_{T_{n}}-U|>\varepsilon)=\int f_{n}\mathrm{d}\mathsf{P}_{U}$
where $f_{n}(u)=P_{W_{T_{n}}}\left(\left|a_{T}(u,W_{T_{n}})-u\right|>\varepsilon\right)$.
Condition \emph{(a)} indicates that $f_{n}\rightarrow0$, and then
by $|f_{n}|\le1$ and the dominated convergence theorem, we have $\int f_{n}\mathrm{d}\mathsf{P}_{U}\rightarrow0$,
which implies that $Z_{T_{n}}\overset{P}{\rightarrow}U$. Moreover,
$Z_{T_{n}}\overset{P}{\rightarrow}U$ implies $(Z_{T_{n}},Z_{\eta})\overset{P}{\rightarrow}(U,Z_{\eta})$,
where $Z_{\eta}=a_{\eta}(U,V_{\eta})$. Then by the continuous mapping
theorem and condition \emph{(b)} we obtain $W_{b_{n}}\overset{P}{\rightarrow}b(U,a_{\eta}(U,V_{\eta}))=V_{\eta}$
and $W_{b_{n}}(\theta^{*})\overset{P}{\rightarrow}V_{\eta}$.

Next we prove that $\mathbb{E}(f(W_{b_{n}})|Z_{T_{n}})\overset{P}{\rightarrow}\mathbb{E}(f(V_{\eta}))$
for any bounded continuous function $f$, where the notation $\mathbb{E}(X|Y)$
stands for the conditional expectation of $X$ given $Y$. The main
tool to prove this result is Theorem 2.1 of \citet{goggin1994}. Let
$Q_{n}$ be a probability measure under which $W_{b_{n}}$ and $Z_{T_{n}}$
are independent, \emph{i.e.}, $Q_{n}((-\infty,w]\times(-\infty,z])=F_{W_{b_{n}}}(w)F_{Z_{T_{n}}}(z)$,
where $F_{W_{b_{n}}}$ and $F_{Z_{T_{n}}}$ are the corresponding
marginal c.d.f.'s. Then for any $\varepsilon>0$, under the $Q_{n}$
measure, $P_{Q_{n}}(|l_{n}(W_{b_{n}},Z_{T_{n}})-1|>\varepsilon)=\int I_{A_{n}}\mathrm{d}Q_{n}$,
where $I_{A_{n}}$ is the indicator function of the set $A_{n}=\{(w,z):|l_{n}(w,z)-1|>\varepsilon\}$.
Condition \emph{(c)} implies that $I_{A_{n}}\rightarrow0$ pointwisely,
so by the dominated convergence theorem we have $\int I_{A_{n}}\mathrm{d}Q_{n}\rightarrow0$.
As a result, under the $Q_{n}$ measure, $l_{n}(W_{b_{n}},Z_{T_{n}})\overset{P}{\rightarrow}1$
and hence $(W_{b_{n}},Z_{T_{n}},l_{n}(W_{b_{n}},Z_{T_{n}}))\overset{d}{\rightarrow}(V_{\eta},U,1)$.
Then Theorem 2.1 of \citet{goggin1994} claims that $\mathbb{E}(f(W_{b_{n}})|Z_{T_{n}})\overset{d}{\rightarrow}\mathbb{E}(f(V_{\eta})|U)$
for any bounded continuous function $f$ . Since $U$ and $V_{\eta}$
are independent, we have $\mathbb{E}(f(V_{\eta})|U)=\mathbb{E}(f(V_{\eta}))$
and hence $\mathbb{E}(f(W_{b_{n}})|Z_{T_{n}})\overset{P}{\rightarrow}\mathbb{E}(f(V_{\eta})$.

Finally, Theorem 2.1 of \citet{XIONG20083249} shows that $\mathbb{E}(f(W_{b_{n}})|Z_{T_{n}})\overset{P}{\rightarrow}\mathbb{E}(f(V_{\eta}))$
is equivalent to $W_{b_{n}}|Z_{T_{n}}\overset{d.P}{\rightarrow}V_{\eta}$,
which concludes the proof.

\subsection{Proof of \eqref{eq:normal_1_cdf}, \eqref{eq:normal_1_cpl}, and
\eqref{eq:normal_1_interval}}

Let $\mathbf{0}_{k}$ denote the $k\times1$ zero vector, $I_{k}$
be the $k\times k$ identity matrix, and $J_{k}$ be a $k\times k$
matrix with all elements being one. It is easy to show that $(W_{b},W_{H}')'=A(e',\varepsilon')'$,
where $A=\left(\begin{array}{cccc}
\frac{1}{n} & \frac{1}{n}\mathbf{1}_{n-1}' & (\frac{1}{n}-1)\tau & \frac{\tau}{n}\mathbf{1}_{n-1}'\\
-\mathbf{1}_{n-1} & I_{n-1} & -\tau\mathbf{1}_{n-1} & \tau I_{n-1}
\end{array}\right)$, $e=(e_{1},\ldots,e_{n})'$, and $\varepsilon=(\varepsilon_{1},\ldots,\varepsilon_{n})'$.
Since $(e',\varepsilon')'\sim\mathsf{N}(\mathbf{0}_{2n},I_{2n})$,
we have $(W_{b},W_{H}')'\sim\mathsf{N}\left(\mathbf{0}_{2n},\left(\begin{smallmatrix}\Sigma_{11} & \Sigma_{12}\\
\Sigma_{21} & \Sigma_{22}
\end{smallmatrix}\right)\right)$, where $\Sigma_{11}=\{1+(n-1)\tau^{2}\}/n$, $\Sigma_{12}=\tau^{2}\mathbf{1}_{n-1}'$,
and $\Sigma_{22}=(\tau^{2}+1)(J_{n-1}+I_{n-1})$.

Simple calculation shows that $\Sigma_{22}^{-1}=(\tau^{2}+1)^{-1}(I_{n-1}-n^{-1}J_{n-1})$,
and then according to the property of multivariate normal distribution,
we have $W_{b}|W_{H}=h\sim\mathsf{N}(\tilde{\mu},\tilde{\sigma}^{2})$,
where $\tilde{\mu}=\Sigma_{12}\Sigma_{22}^{-1}h=\tau^{2}(\tau^{2}+1)^{-1}(\bar{x}-x_{1})$,
and $\tilde{\sigma}^{2}=\Sigma_{11}-\Sigma_{12}\Sigma_{22}^{-1}\Sigma_{21}=n^{-1}(1+\tau^{2})^{-1}(n\tau^{2}+1)$.

Let $F_{W_{b}|h}$ denote the c.d.f. of $\mathsf{N}(\tilde{\mu},\tilde{\sigma}^{2})$,
then
\begin{align*}
\mathcal{S}_{h} & =\left\{ F_{W_{b}|h}^{-1}(u'),u'\in(0,1):\vert u'-0.5\vert<\vert U_{\mathcal{S}}-0.5\vert\right\} ,\ U_{\mathcal{S}}\sim\mathsf{Unif}(0,1)\\
 & =\left\{ z:\vert(z-\tilde{\mu})/\tilde{\sigma}\vert<\vert Z_{\mathcal{S}}\vert\right\} ,\ Z_{\mathcal{S}}\sim\mathsf{N}(0,1).
\end{align*}
Therefore, define $Q_{h}(s)=P_{\mathcal{S}_{h}}(s\notin\mathcal{S}_{h})$,
and we get $Q_{h}(s)=2\Phi(|(s-\tilde{\mu})/\tilde{\sigma}|)-1$.
From \eqref{eq:cpl_Q} we have $\mathsf{cpl}_{T(x)|h}(\mu_{1})=1-Q_{h}(w)$
where $w=b(T(x),\mu_{1})=\bar{x}-\mu_{1}$. As a result, $\mathsf{cpl}_{T(x)|h}(\mu_{1})=2-2\Phi(|(\bar{x}-\mu_{1}-\tilde{\mu})/\tilde{\sigma}|)=2\Phi\left(-|(\bar{x}-\mu_{1}-\tilde{\mu})/\tilde{\sigma}|\right)$,
which reduces to \eqref{eq:normal_1_cpl}. The interval estimator
then follows directly.

\subsection{Proof of \eqref{eq:normal_2_mu_sigma} and \eqref{eq:normal_2_cdf}}

Let $U=\left(\tau\varepsilon_{1}+e_{1}-\tilde{\tau}Z\right)/(\sqrt{n}\tilde{\tau})$,
and then it is easy to verify that $(U,e_{1})'\sim\mathsf{N}(\mathbf{0},\Sigma)$,
where $\Sigma=\left(\begin{smallmatrix}1 & \rho\\
\rho & 1
\end{smallmatrix}\right)$ and $\rho=(\sqrt{n}\tilde{\tau})^{-1}$. Since $U$, $e_{1}$, and
$M_{n-2}^{2}$ are independent, the joint density function of $(U,e_{1},M_{n-2}^{2})$
can be written as
\[
g_{0}(u,z,x)\propto\exp\left\{ -\frac{1}{2}(u,z)\Sigma^{-1}(u,z)'\right\} x^{\frac{n}{2}-2}\exp\left\{ -\frac{n-2}{2}x\right\} .
\]
Let $W_{e}=e_{1}/\sqrt{M_{n-2}^{2}}$. Note also that $W_{H}=U/\sqrt{M_{n-2}^{2}}$,
so with the transformation of variables $s=x/\sqrt{z},h=y/\sqrt{z},t=z$,
the joint density of $(W_{e},W_{H},M_{n-2}^{2})$ is
\begin{align*}
g(s,h,t) & \propto\exp\left\{ -\frac{1}{2}(s\sqrt{t},h\sqrt{t})\Sigma^{-1}(s\sqrt{t},h\sqrt{t})'\right\} t^{\frac{n}{2}-2}\exp\left\{ -\frac{n-2}{2}t\right\} \cdot t\\
 & =\exp\left\{ -\frac{1}{2}t(s,h)\Sigma^{-1}(s,h)'\right\} t^{\frac{n}{2}-1}\exp\left\{ -\frac{n-2}{2}t\right\} .
\end{align*}
For simplicity of notations let $\Sigma^{-1}=\left(\begin{smallmatrix}A & B\\
B & A
\end{smallmatrix}\right)$, where $A=n(\tau^{2}+1)(n\tau^{2}+1)^{-1},B=-(n\tau^{2}+1)^{-1}\sqrt{n(n-1)(\tau^{2}+1)}$,
and then the joint density of $(W_{e},M_{n-2}^{2})$ given $W_{H}=h$
is
\begin{align}
g(s,t|h) & \propto\exp\left\{ -\frac{1}{2}t(As^{2}+2Bsh+Ah^{2})\right\} t^{\frac{n}{2}-1}\exp\left\{ -\frac{n-2}{2}t\right\} \nonumber \\
 & =\exp\left\{ -\frac{A}{2}t\left(s+\frac{B}{A}h\right)^{2}\right\} \cdot t^{\frac{n}{2}-1}\cdot\exp\left\{ -\frac{1}{2}(h^{2}+n-2)t\right\} .\label{eq:density_x_z_given_y}
\end{align}
Integrating $s$ out gives $g(t|h)\propto t^{(n-1)/2-1}\exp\left\{ -(h^{2}+n-2)t/2\right\} $,
which corresponds to the $2(h^{2}+n-2)^{-1}\mathsf{Gamma}((n-1)/2)$
distribution. \eqref{eq:density_x_z_given_y} also shows that given
$W_{H}=h$ and $M_{n-2}^{2}=t$, the density function of $W_{e}$
is $g(s|h,t)\propto\exp\left\{ -At(s+hB/A)^{2}/2\right\} $, implying
the $\mathsf{N}\left(-hB/A,(At)^{-1}\right)$ distribution.

As a consequence, given $W_{H}=h$, the random variables $M_{n-2}^{2}$
and $W_{e}$ can be expressed as $M_{n-2}^{2}=C\tilde{M}^{2}$ and
$W_{e}=-hB/A+(AM_{n-2}^{2})^{-1/2}\tilde{Z}$, where $C=(n-1)(h^{2}+n-2)^{-1}$,
$\tilde{M}^{2}\sim\chi_{n-1}^{2}/(n-1)$, $\tilde{Z}\sim\mathsf{N}(0,1)$,
and $\tilde{M}^{2}$ and $\tilde{Z}$ are independent. Therefore,
$e_{1}=W_{e}\sqrt{M_{n-2}^{2}}=h\sqrt{\omega(n-1)/n}\cdot\sqrt{C}\tilde{M}+\sqrt{1-\omega(n-1)/n}\cdot\tilde{Z}$.

Now consider the distribution of $\overline{X}_{(-1)}-\mu_{1}$. It
is easy to see that $\overline{X}_{(-1)}-\mu_{1}=e_{1}-\sqrt{n}\tilde{\tau}W_{H}M_{n-2}$,
so given $W_{H}=h$,
\begin{align*}
\mathbb{E}(\overline{X}_{(-1)}-\mu_{1}|W_{H}=h) & =\mathbb{E}(e_{1}|W_{H}=h)-\sqrt{n}\tilde{\tau}h\mathbb{E}(M_{n-2}|W_{H}=h)\\
 & =h\sqrt{C}\left(\sqrt{\omega(n-1)/n}-1/\sqrt{\omega(n-1)/n}\right)\mathbb{E}(\tilde{M}).
\end{align*}
Also $\mathbb{E}(S_{(-1)}|W_{H}=h)=\sqrt{C/\omega}\cdot\mathbb{E}(\tilde{M})$,
$\mathbb{E}(S_{(-1)}^{-1}|W_{H}=h)=\sqrt{\omega/C}\cdot\mathbb{E}(\tilde{M}^{-1})=(n-1)(n-2)^{-1}\sqrt{\omega/C}\cdot\mathbb{E}(\tilde{M})$,
so with the $\tilde{\mu}$ given in \eqref{eq:normal_2_mu_sigma},
we can show that $\mathbb{E}(\overline{X}_{(-1)}-\mu_{1}-\tilde{\mu}|W_{H}=h)=0$.
Similarly, it can be calculated that
\[
\mathrm{Var}(\overline{X}_{(-1)}-\mu_{1}-\tilde{\mu}|W_{H}=h)=1-\frac{(n-1)(n-2)(n-3-h^{2})}{n(n-3)(n-2+h^{2})}\omega,
\]
and an unbiased and consistent estimator for $\omega$ is $\hat{\omega}=(n-3)(h^{2}+n-2)^{-1}S_{(-1)}^{-2}$.
Therefore, with the $\tilde{\sigma}$ in \eqref{eq:normal_2_mu_sigma},
$W_{b}(\tau)|W_{H}=h\overset{d}{\rightarrow}\mathsf{N}(0,1)$ for
any $\tau>0$. The $n^{-\gamma}$ term is used to guarantee that the
variance is always positive.

Finally, the auxiliary variable to predict is
\[
W_{b}(\tau)=\frac{c_{2}\sqrt{\omega}\left(\tilde{M}-c_{3}\tilde{M}^{-1}\right)+\sqrt{1-\omega(n-1)/n}\tilde{Z}}{\sqrt{\max\left\{ n^{-\gamma},1-c_{1}\omega\tilde{M}^{-2}\right\} }},
\]
and \eqref{eq:normal_2_cdf} follows immediately.

\subsection{Computation for the Poisson Hierarchical Model}

\label{subsec:poisson_computation}

We first obtain the expression for $\ell(\lambda_{1};x)$. Given $\lambda_{1}$,
$X_{1}\sim\mathsf{Pois}(\lambda_{1}t_{1})$, $X_{i}=F_{\lambda_{1}t_{i}V_{i}/V_{1}}^{-1}(U_{i})$,
and $X_{1}$ and $X_{(-1)}=(X_{2},\ldots,X_{n})'$ are independent.
Marginally $X_{i}$ follows a negative binomial distribution $\mathsf{NB}(s,p)$
with m.g.f. $p(x)\propto p^{s}(1-p)^{x}$, where $p=1/(1+\gamma)$.
Therefore, the joint density of $X_{(-1)}$ and $V_{1}$ is
\[
p(x_{2},\ldots,x_{n},v_{1}|\lambda_{1})=\prod_{i=2}^{n}\left\{ \frac{\Gamma(x_{i}+s)}{\Gamma(s)}p_{i}^{s}(1-p_{i})^{x_{i}}\right\} \cdot\frac{1}{\Gamma(s)}v_{1}^{s-1}e^{-v_{1}},\quad p_{i}=\frac{1}{1+\lambda_{1}t_{i}/v_{1}},
\]
and hence the density of $X_{(-1)}$ is
\[
p(x_{2},\ldots,x_{n}|\lambda_{1})=\int_{0}^{+\infty}\prod_{i=2}^{n}\left\{ \frac{\Gamma(x_{i}+s)}{\Gamma(s)}p_{i}^{s}(1-p_{i})^{x_{i}}\right\} \cdot\frac{1}{\Gamma(s)}v_{1}^{s-1}e^{-v_{1}}\mathrm{d}v_{1}.
\]
As a result, $\ell(\lambda_{1};x)=x_{1}\log(\lambda_{1})-\lambda_{1}t_{1}+\log p(x_{2},\ldots,x_{n}|\lambda_{1})+C$,
where $C$ is some constant unrelated to $\lambda_{1}$, and the MLE
for $\lambda_{1}$ can be obtained using standard optimization methods.

To obtain $G_{\lambda_{1}}$, the c.d.f. of $W_{b}(\lambda_{1})$,
we first use Monte Carlo method to simulate $U$ and $V$ to get a
random sample of $W_{b}(\lambda_{1})$, and then $G_{\lambda_{1}}$
is approximated by $\hat{G}_{\lambda_{1}}$, the empirical c.d.f.
of $W_{b}(\lambda_{1})$. Finally, the interval estimator is computed
using a grid search on $\mathsf{pl}_{x}(\lambda_{1})$.

\subsection{Computation for the Binomial Rates-Difference Model}

\label{subsec:binomial_computation}

It is easy to show that $\ell(\delta,\omega;x,y)=x\log p_{1}+(m-x)\log(1-p_{1})+y\log p_{2}+(n-y)\log(1-p_{2})+\log\pi(\delta),$
where $p_{1}=\{1+\delta+(1-|\delta|)\omega\}/2$ and $p_{2}=\{1-\delta+(1-|\delta|)\omega\}/2$.

Keeping $\delta$ fixed, $\hat{\omega}_{\delta}=\arg\max_{\omega}\ell(\delta,\omega;x,y)$
can be obtained by solving the equation
\begin{equation}
\frac{\partial\ell}{\partial\omega}=\left(\frac{x}{p_{1}}-\frac{m-x}{1-p_{1}}+\frac{y}{p_{2}}-\frac{n-y}{1-p_{2}}\right)\cdot\frac{1}{2}(1-|\delta|)=0.\label{eq:binomial_equation_w}
\end{equation}
Since $p_{1}=p_{2}+\delta$, \eqref{eq:binomial_equation_w} reduces
to a cubic equation $ap_{2}^{3}+bp_{2}^{2}+cp_{2}+d=0$, where $a=m+n$,
$b=-(x+y)-m(1-\delta)-n(1-2\delta)$, $c=x-m\delta+y(1-2\delta)-n(\delta-\delta^{2})$,
and $d=y(\delta-\delta^{2})$. The solution should be sought within
the range $\max(0,-\delta)<p_{2}<\min(1,1-\delta)$. As a result,
$(\hat{\delta},\hat{\omega})=\arg\max_{\delta,\omega}\ell(\delta,\omega;x,y)$
is obtained by computing $\hat{\omega}_{\delta}$ over a grid of $\delta$
values.

The remaining part of the computation proceeds similarly to the Poisson
model, by simulating $(U_{1},U_{2},U)$ and computing the distribution
of $W_{b}(\omega)$, and hence the details are omitted.

\bibliographystyle{asa}
\nocite{*}
\bibliography{ref}

\end{document}